\documentclass[lettersize,journal]{IEEEtran}
\usepackage{amsmath,amsfonts}
\usepackage{algorithm}
\usepackage{xcolor}
\usepackage[table]{xcolor}
\definecolor{darkgreen}{RGB}{0,180,0} 

\usepackage{array}
\usepackage[caption=false,font=normalsize,labelfont=sf,textfont=sf]{subfig}
\usepackage{textcomp}
\usepackage{stfloats}
\usepackage{url}
\usepackage{verbatim}
\usepackage{cite}

\newcommand{\LL}[1]{\textcolor{black}{#1}}
\usepackage[normalem]{ulem}
\usepackage{algpseudocode}
\usepackage{multirow}
\usepackage{threeparttable}
\usepackage{graphicx}
\usepackage{enumitem}
\usepackage{booktabs}

\begin{document}

\title{IGDMRec: Behavior Conditioned Item Graph Diffusion for Multimodal Recommendation}

\author{Ziyuan Guo, Jie Guo,~\IEEEmembership{Senior Member,~IEEE,}
Zhenghao Chen, Bin Song,~\IEEEmembership{Senior Member,~IEEE},\\ and Fei Richard Yu,~\IEEEmembership{Fellow,~IEEE}
\thanks{Ziyuan Guo, Jie Guo, and Bin Song are with the State Key Laboratory of Integrated Services Networks, Xidian University, Xi’an, Shaanxi 710071, China (e-mail: ziyuanguo@stu.xidian.edu.cn; jguo@xidian.edu.cn; bsong@mail.xidian.edu.cn).\\ \hspace*{1em}Zhenghao Chen is with the Hangzhou Institute of Technology, Xidian University, Hangzhou 311231, China (e-mail: 24241214851@stu.xidian.edu.cn).\\ \hspace*{1em}Fei Richard Yu is with the School of Information Technology, Carleton University, Canada (e-mail: richard.yu@carleton.ca).\\ \hspace*{1em}Corresponding author: Jie Guo.}
}

\markboth{Journal of \LaTeX\ Class Files,~Vol.~14, No.~8, August~2021}%
{Shell \MakeLowercase{\textit{et al.}}: A Sample Article Using IEEEtran.cls for IEEE Journals}


\maketitle

\begin{abstract}
Multimodal recommender systems (MRSs) are critical for various online platforms, offering users more accurate personalized recommendations by incorporating multimodal information of items. Structure-based MRSs have achieved state-of-the-art performance by constructing semantic item graphs, which explicitly model relationships between items based on modality feature similarity. However, such semantic item graphs are often noisy due to 1) inherent noise in multimodal information and 2) misalignment between item semantics and user-item co-occurrence relationships, which introduces false links and leads to suboptimal recommendations. To address this challenge, we propose {I}tem {G}raph {D}iffusion for {M}ultimodal {Rec}ommendation (IGDMRec), a novel method that leverages a diffusion model with classifier-free guidance to denoise the semantic item graph by integrating user behavioral information. Specifically, IGDMRec introduces a Behavior-conditioned Graph Diffusion (BGD) module, incorporating interaction data as conditioning information to guide the denoising of the semantic item graph. Additionally, a Conditional Denoising Network (CD-Net) is designed to implement the denoising process with manageable complexity. Finally, we propose a contrastive representation augmentation scheme that leverages both the denoised item graph and the original item graph to enhance item representations. \LL{Extensive experiments on four real-world datasets demonstrate the superiority of IGDMRec over competitive baselines, with robustness analysis validating its denoising capability and ablation studies verifying the effectiveness of its key components.} 
\end{abstract}


\section{Introduction}
\IEEEPARstart{R}{ecommender} systems (RSs) have emerged as essential tools for assisting users in finding content of interest from vast item collections~\cite{RS-1,RS-2,RS-3,RS-4}. General RSs focus on modeling user preferences through interaction data (e.g., clicks, ratings, purchase records), serving as the cornerstone of personalized recommendations~\cite{CF-1,CF-2}. In recent years, multimodal recommender systems (MRSs) have garnered widespread attention for enhancing recommendation accuracy by incorporating rich multimodal information of items (e.g., text, images, videos) as auxiliary information~\cite{SIGIR23,hypergraph,MGCRec}.

Conventional MRSs enhance item representations by directly integrating multimodal features~\cite{VBPR,ACF}, \LL{while recent methods leverage graph neural networks (GNNs) to propagate multimodal information over the user–item interaction graph to better model user preferences~\cite{MMGCN,DualGNN,GRCN}.} However, these feature-based methods model item relationships implicitly through higher-order item-user-item co-occurrence, ignoring the intrinsic multimodal features of items~\cite{LATTICE}. \LL{Structure-based MRSs address this by explicitly constructing semantic item graphs from modality feature similarity to mine latent item structures, showing clear performance gains~\cite{MVGC,structure-1,LATTICE,freedom,KDD24,SPACE}.}

\LL{Nevertheless, structure-based MRSs typically construct the semantic item graph by evaluating the similarity of modality features between items, resulting in a critical issue: \textbf{the semantic item graph is noisy}. This noise arises from two intertwined factors:
\begin{itemize}
    \item \textit{Modality-inherent noise.} Multimodal features often contain irrelevant or misleading details~\cite{MVGC,KDD24,MICRO}, such as redundant text descriptions or image backgrounds, which distort semantic similarity estimation.
    \item \textit{Semantic–behavioral misalignment.} User preferences are driven by diverse and complex factors~\cite{pre-1,pre-2}, meaning that semantically similar items do not necessarily attract the same users, while semantically dissimilar items may still exhibit strong co-preference.
\end{itemize}
These two factors jointly lead to false-positive and false-negative links in the semantic item graph that are misaligned with user preferences. A \textit{false-positive link} occurs when similar items are not jointly preferred (e.g., sporty and fashionable backpacks that share similar materials or appearances but appeal to different user groups), while a \textit{false-negative link} arises when semantically dissimilar items are behaviorally related but remain unconnected (e.g., two visually distinct backpacks that may be frequently co-purchased). Recently, self-supervised learning (SSL) methods have been explored to alleviate semantic noise by constructing supervision signals from augmented views~\cite{MMGCL,BM3,SLMRec}. However, most SSL-based MRSs rely on manually designed perturbations (e.g., masking, dropout), which may fail to produce high-quality views for reliable optimization. Therefore, effectively leveraging interaction data that directly reflects user preferences to explicitly denoise the semantic item graph is crucial for MRSs and is not adequately addressed by existing methods.
}

Recent studies have leveraged advanced diffusion models (DMs)~\cite{DDPM,Diffbeats} to denoise the user-item interaction graph, learning to reconstruct original interactions from corrupted ones for recommendation tasks~\cite{DIFFREC,CF-Diff,GIFFCF}. The potential advantages of DMs in graph structure denoising are twofold: 
1) \LL{The forward–reverse diffusion paradigm aligns with the denoising objective, as it learns to reconstruct high quality samples from corrupted ones through iterative refinement.}
2) DMs decompose the denoising task into multiple iterative steps, which allows iterative updates of the graph structure toward fine-grained directions. 
Thus, it is promising to leverage DMs to denoise the semantic item graph. Importantly, to enhance item relationships consistent with users' preferences, we propose integrating interaction data into the diffusion process of the item graph, thereby generating a denoised item graph that incorporates both semantic and behavioral information. However, improving recommendation through DM-based item graph denoising presents the following two challenges:
\begin{itemize}
    \item \textbf{C1}: How to efficiently integrate behavioral information into the item graph using DMs to achieve denoising.
    \item \textbf{C2}: How to fully utilize both the denoised and original item graphs for accurate recommendations.
\end{itemize}

To address the above challenges, we propose a novel Item Graph Diffusion for Multimodal Recommendation (IGDMRec). Drawing inspiration from conditional DMs~\cite{class-free}, our method leverages the controlled generation capabilities of conditional DMs to generate denoised diffusion-aware item graphs and enable accurate recommendations through contrastive learning.
First, IGDMRec constructs a behavioral item graph based on item-user-item co-occurrence relationships in interactions. To efficiently integrate behavioral information into the semantic item graph, a Behavior-conditioned Graph Diffusion (BGD) module is proposed (solving the challenge \textbf{C1}). Specifically, the BGD module gradually corrupts semantic relationships by injecting Gaussian noise in the forward process and iteratively recovers them through the reverse process conditioned on the behavioral item graph.
Meanwhile, a conditional denoising network (CD-Net) is proposed to implement the reverse process with manageable complexity. Finally, IGDMRec introduces a contrastive representation augmentation scheme that enables the full utilization of both the diffusion-aware and the semantic item graphs to improve recommendation performance (solving the challenge \textbf{C2}). The main contributions of this paper are summarized as follows:

\begin{itemize}
    \item We propose a novel multimodal recommendation method, IGDMRec, which models the structure optimization of the item graph as a diffusion process and employs classifier-free guidance to enhance the alignment between the generated item graph and user behavioral information.
    \item A BGD module with a lightweight CD-Net is designed to denoise the item graph by integrating behavioral information into the diffusion process. Additionally, contrastive learning is introduced to enhance item representations.
    \item \LL{Extensive experiments on four datasets validate the effectiveness of the proposed IGDMRec and its substantial performance improvement over the baselines.} The ablation study confirms the contribution of each component.
\end{itemize}

\section{Related Work}
\LL{In this section, we present recent work related to MRSs, followed by diffusion models for recommendation.}
\subsection{Multimodal Recommendation}
MRSs aim to enhance recommendations by incorporating multimodal information of items. \textit{Feature-based MRSs} directly utilize multimodal features to enhance item representations. VBPR~\cite{VBPR} concatenates multimodal features with latent item representations obtained by matrix factorization, and ACF~\cite{ACF} employs an attention mechanism that leverages multimodal features to capture component-level user preferences. Recent studies, such as MMGCN~\cite{MMGCN} and GRCN~\cite{GRCN}, leverage GCNs to model higher-order user-item relationships in multimodal scenarios. To further capture implicit item representations, \textit{structure-based MRSs} have gained significant attention and achieved state-of-the-art performance by mining latent item structures. LATTICE~\cite{LATTICE} constructs a dynamic semantic item graph for each modality to enhance item representations, while FREEDOM~\cite{freedom} further freezes the semantic item graph and uses a degree-sensitive edge pruning method to denoise the interaction graph. However, most existing structure-based MRSs overlook the structural optimization of item graphs. The presence of noise in semantic item graphs hinders effective item representation learning, ultimately degrading recommendation performance.

\LL{Self-supervised learning (SSL) has been extensively investigated in recommender systems for its ability to generate supervision signals from inherent data structures, thereby mitigating the issue of data sparsity. Contrastive learning, which enhances representation learning by maximizing consistency across different views, has become the predominant paradigm in MRSs. Early \textit{data-based} methods constructe views directly from raw data. For instance, MMGCL~\cite{MMGCL}  employs modality masking and edge dropout to create augmented interaction graph views, while SLMRec~\cite{SLMRec} adopts feature dropout and masking to generate two augmented views for each item. As for the \textit{feature-based} methods, BM3~\cite{BM3} encodes modality embeddings, applies embedding dropout to construct augmented views, and introduces an ID-based embedding view for multimodal alignment. Recently, MICRO~\cite{MICRO}, as a representative \textit{model-based} method, constructs modality-specific semantic item graphs and employs GNNs to encode both modality-specific and fused views for contrastive learning.}

\LL{While SSL inherently exhibits robustness to semantic noise, existing SSL-based MRSs typically rely on manually designed view construction strategies (e.g., masking, dropout), which face challenges in creating high-quality views for optimization. Furthermore, existing methods lack an explicit mechanism to denoise the semantic item graph, while mining high-quality item relationships has been shown to be crucial for recommendations~\cite{LATTICE,freedom}. In contrast, IGDMRec explicitly denoises the semantic item graph. Unlike prior random augmentation strategies, IGDMRec generates a high-quality view through a behavior-conditioned graph denoising process and further refines item representations through contrastive learning. Furthermore, IGDMRec can explicitly correct false-positive and false-negative links in the semantic item graph, thereby mining more accurate latent item structures.}


\subsection{Diffusion Models for Recommendation}
Diffusion models (DMs), as a novel generative paradigm, have garnered widespread attention across diverse domains~\cite{Diffbeats,DiffTrans}.
Recently, several studies have introduced DMs into recommendations. For instance, CODIGEM~\cite{CODIGEM} and DiffRec~\cite{DIFFREC} leverage DMs to model the distribution of discrete user-item interactions, thereby inferring user preferences. Beyond general RSs, DMs have also been used in knowledge graph-based recommendation~\cite{DiffKG}, social recommendation~\cite{RecDiff}, and sequential recommendation~\cite{DiffuRec,Genearate-what}. In the context of MRSs, MCDRec~\cite{MCDRec} uses DMs to integrate multimodal features directly into item representations and denoise the user-item graph via diffusion-aware knowledge. DiffMM~\cite{DiffMM} performs a diffusion process on the user-item graph while injecting multimodal information to generate modality-aware user-item graphs, and enhances recommendations with self-supervised learning. 

Despite the impressive performance of current diffusion-based RSs and MRSs, existing works overlook the modeling and optimization of the structure between items. However, explicitly modeling high-quality relationships between items is crucial for exploring implicit item representations, which have been shown to significantly enhance recommendations~\cite{MVGC,structure-1,LATTICE,freedom,KDD24}. 
In our proposed IGDMRec, we model the structure optimization of the semantic item graph as a diffusion process and integrate behavioral information through classifier-free guidance, thereby enhancing item representations and ultimately improving recommendations. 

\section{Preliminary}
In this section, we present the conditional diffusion model based on the diffusion probabilistic model framework~\cite{DPM,understand_diffusion}. Then, the notations of this paper are given.

\subsection{Conditional Diffusion Model}
\label{Sec:CDM}
\subsubsection{Forward (Noising) Process} Given an input data $\boldsymbol{x}_0\sim q(\boldsymbol{x}_0)$, the forward process gradually perturbs the original data $\boldsymbol{x}_0$ by adding Gaussian noise: for $t=1,\dots T$,
\begin{equation}
\label{eqn:Diff-Forward}
    q(\boldsymbol{x}_t|\boldsymbol{x}_{t-1})=\mathcal{N}(\boldsymbol{x}_t;\sqrt{1-\beta_t}\boldsymbol{x}_{t-1},\beta_t\boldsymbol{I}),
\end{equation}
where $\mathcal{N}(\cdot)$ denotes the Gaussian distribution, and $\beta_t\in(0,1)$ controls the noise scale at the $t$-th step. 
By applying the reparameterization trick~\cite{DDPM}, we can obtain $\boldsymbol{x}_t$ directly from $\boldsymbol{x}_0$ as follows:
\begin{equation}
    \label{Eqn:x_t-x_0}
    \boldsymbol{x}_t=\sqrt{\bar{\alpha}_t}\boldsymbol{x}_{0}+\sqrt{1-\bar{\alpha}_t}\boldsymbol{\epsilon},\ \  \boldsymbol{\epsilon}\sim\mathcal{N}(\boldsymbol{0},\boldsymbol{I}),
\end{equation}
where $\bar{\alpha}_t=\textstyle{\prod}_{s=1}^t\alpha_s$, $\alpha_s=1-\beta_s$, and $[\beta_1,\dots,\beta_T]$ is the predefined noise schedule used to control step-wise noise.

\subsubsection{Reverse (Denoising) Process}
The reverse process of the conditional DM gradually reconstructs the original data $\boldsymbol{x}_0$ from $\boldsymbol{x}_T$ by an iterative approach: for $t=T,\dots,1$,
\begin{equation}
\label{Eqn:reverse}
    p_{\theta}(\boldsymbol{x}_{t-1}|\boldsymbol{x}_t,\boldsymbol{c})=\mathcal{N}(\boldsymbol{x}_t;\boldsymbol{\mu}_{\theta}(\boldsymbol{x}_t,\boldsymbol{c},t),\boldsymbol{\Sigma}_{\theta}(\boldsymbol{x}_t,\boldsymbol{c},t)),
\end{equation}
where $\boldsymbol{c}$ denotes the conditioning information of the reverse process, $\boldsymbol{\mu}_{\theta}(\boldsymbol{x}_t,\boldsymbol{c},t)$ and $\boldsymbol{\Sigma}_{\theta}(\boldsymbol{x}_t,\boldsymbol{c},t)$ are the mean and variance learned by the neural network parameterized by $\theta$, respectively. 
Importantly, based on Eq.~\eqref{eqn:Diff-Forward} and Eq.~\eqref{Eqn:x_t-x_0}, the reverse probability conditional on $\boldsymbol{x}_0$ can be resolved using the Bayes rule:
\begin{equation}
\label{Eqn:reverse-bayes}
    q(\boldsymbol{x}_{t-1}|\boldsymbol{x}_t,\boldsymbol{x}_0)\propto\mathcal{N}(\boldsymbol{x}_{t-1};\tilde{\boldsymbol{\mu}}_t(\boldsymbol{x}_t,\boldsymbol{x}_0,t),\tilde{\beta}_t\boldsymbol{I}),
\end{equation}
where
\begin{equation}
    \begin{aligned}
        \label{Eqn:bayes-mv}\tilde{\boldsymbol{\mu}}_t(\boldsymbol{x}_t,\boldsymbol{x}_0,t)&=\frac{\sqrt{\alpha_t}(1-\bar{\alpha}_{t-1})}{1-\bar{\alpha}_t}\boldsymbol{x}_t+\frac{\sqrt{\bar{\alpha}_{t-1}}(1-{\alpha}_{t})}{1-\bar{\alpha}_t}\boldsymbol{x}_0,\;\text{and}\\
        \tilde{\beta}_t &= \frac{(1-\alpha_t)(1-\bar{\alpha}_{t-1})}{1-\bar{\alpha}_t}.
    \end{aligned}
\end{equation}

\subsubsection{Optimization} With the aim of generating the original data $\boldsymbol{x}_0$, the conditional DM is optimized by maximizing the evidence lower bound (ELBO) of $\log p(\boldsymbol{x}_0)$:
\begin{equation}
    \log p(\boldsymbol{x}_0)\geq\mathbb{E}_{q(\boldsymbol{x}_{1:T}|\boldsymbol{x}_0)}\left[\log\frac{p_{\theta}(\boldsymbol{x}_{0:T})}{q(\boldsymbol{x}_{1:T}|\boldsymbol{x}_0)}\right]:=\text{ELBO}\ .
\end{equation}
As outlined in~\cite{understand_diffusion}, the optimization objective can be expressed as $\textstyle{\sum}_{t=1}^T\mathcal{L}_t$, with each component defined as:
\begin{small}
    \begin{equation}
    \label{Eqn:ELBO}
        \mathcal{L}_{t} \!\!= \!\!\left\{
        \begin{aligned}
        &\mathbb{E}_{q(\boldsymbol{x}_{1}|\boldsymbol{x}_0)} \!
        \left[-\log p_{\theta}(\boldsymbol{x}_0|\boldsymbol{x}_1)\right], & t=1, \\
        &\mathbb{E}_{q(\boldsymbol{x}_{t}|\boldsymbol{x}_0)} \!
        \left[ D\left(q(\boldsymbol{x}_{t-1}|\boldsymbol{x}_t,\boldsymbol{x}_0) 
        \middle\| p_{\theta}(\boldsymbol{x}_{t-1}|\boldsymbol{x}_t,\boldsymbol{c})\right) \right],\!\!\! & t \geq 2.
        \end{aligned}
        \right.
    \end{equation}
\end{small}

\noindent where $D(\cdot||\cdot)$ denotes the KL divergence between two distributions. For $t=1$, the loss $\mathcal{L}_1$ can be simplified as:
\begin{equation}
    \label{Eqn:diffL1}\mathcal{L}_1=\mathbb{E}_{q(\boldsymbol{x}_1|\boldsymbol{x}_0)}\big[\|\boldsymbol{x}_0-\hat{\boldsymbol{x}}_{\theta}(\boldsymbol{x}_1,\boldsymbol{c},1)\|_2^2\big],
\end{equation}
where $\hat{\boldsymbol{x}}_{\theta}(\boldsymbol{x}_1,\boldsymbol{c},1)$ is the predicted $\boldsymbol{x}_0$ at time step $t=1$. For $t\geq2$, by substituting Eq.~\eqref{Eqn:reverse} and Eq.~\eqref{Eqn:reverse-bayes} into Eq.~\eqref{Eqn:ELBO}, and setting $\Sigma_{\theta}(\boldsymbol{x}_t,\boldsymbol{c},t)=\tilde{\beta}_t\boldsymbol{I}$ for stable training, we have
\begin{equation}
\label{Eqn:diffLt}
    \mathcal{L}_t\!\! = \!\!\mathbb{E}_{q(\boldsymbol{x}_t|\boldsymbol{x}_0)}\!\Big[\tfrac{1}{2\tilde{\beta}_t}\|\boldsymbol{\mu}_{\theta}(\boldsymbol{x}_t,\boldsymbol{c},t)-\tilde{\boldsymbol{\mu}}_t(\boldsymbol{x}_t,\boldsymbol{x}_0,t)\|_2^2\Big], \ t\geq2.
\end{equation}
We can model $\boldsymbol{\mu}_{\theta}(\boldsymbol{x}_t,\boldsymbol{c},t)$ in a similar form of Eq.~\eqref{Eqn:bayes-mv}: 
\begin{small}
    \begin{equation}
    \label{Eqn:mu}
        \boldsymbol{\mu}_{\theta}(\boldsymbol{x}_t,\boldsymbol{c},t)\!=\!\frac{\sqrt{\alpha_t}(1-\bar{\alpha}_{t-1})}{1-\bar{\alpha}_t}\boldsymbol{x}_t\!+\!\frac{\sqrt{\bar{\alpha}_{t-1}}(1-\alpha_t)}{1-\bar{\alpha}_t}\hat{\boldsymbol{x}}_{\theta}(\boldsymbol{x}_t,\boldsymbol{c},t),
\end{equation}
\end{small}

\noindent where $\hat{\boldsymbol{x}}_{\theta}(\boldsymbol{x}_t,\boldsymbol{c},t)$ denotes the predicted $\boldsymbol{x}_0$ at time step $t$. Thus, by substituting Eq.~\eqref{Eqn:bayes-mv} and Eq.~\eqref{Eqn:mu} into Eq.~\eqref{Eqn:diffLt}, we have: For $t\geq 2$,
\begin{small}
\begin{equation}
    \label{Eqn:condition-train}
    \mathcal{L}_t = \mathbb{E}_{q(\boldsymbol{x}_t|\boldsymbol{x}_0)}\Big[\tfrac{1}{2}\left(\tfrac{\bar{\alpha}_{t-1}}{1-\bar{\alpha}_{t-1}}-\tfrac{\bar{\alpha}_t}{1-\bar{\alpha}_t}\right)\|\boldsymbol{x}_0-\hat{\boldsymbol{x}}_{\theta}(\boldsymbol{x}_t,\boldsymbol{c},t)\|_2^2\Big].
\end{equation}
\end{small}

With the above simplifications, the optimization objective is transformed into predicting $\boldsymbol{x}_0$ by neural networks at each time step. The prediction $\hat{\boldsymbol{x}}_{\theta}(\boldsymbol{x}_t,\boldsymbol{c},t)$ is commonly achieved using networks such as U-Net~\cite{DDPM} and Transformer\cite{DiffTrans}.

\textit{$\bullet$ Note:} Leveraging the strength of the conditional DM in generating high-quality data guided by conditioning information~\cite{DDPM,Diffbeats}, this paper introduces it into MRSs by applying a diffusion process to the semantic item graph. However, a conditional DM trained based on Eq.~\eqref{Eqn:diffL1} and Eq.~\eqref{Eqn:condition-train} may potentially ignore or downplay the provided conditioning information~\cite{understand_diffusion}, which can degrade the quality of the item graph generated by the DM, thereby impairing recommendation performance. To mitigate this problem, we adopt the classifier-free guidance scheme~\cite{class-free} to jointly train both conditional and unconditional DMs to improve recommendation performance, as detailed in Section \ref{BGD}. \LL{Compared with variational autoencoders (VAEs) that rely on a fixed Gaussian latent prior and single-step reconstruction~\cite{vae}, diffusion models refine data generation through iterative denoising with flexible conditional guidance, making them more suitable for explicitly denoising item graphs.}

\begin{figure*}[t!]
  \centering
  \begin{minipage}[t]{0.55\textwidth}
    \centering
    \includegraphics[height=6.35cm]{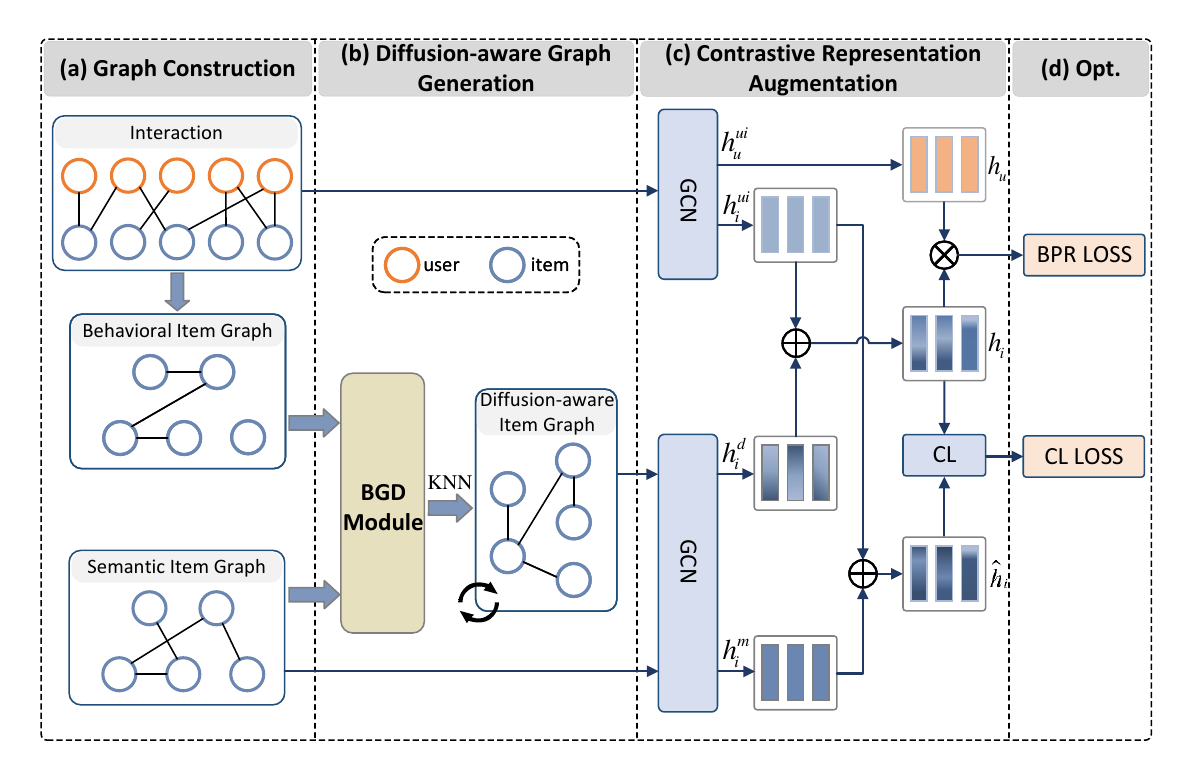}
    \textsf{\scriptsize (a) Overall structure of IGDMRec.}
  \end{minipage}
  \begin{minipage}[t]{0.42\textwidth}
    \centering
    \includegraphics[height=6.31cm]{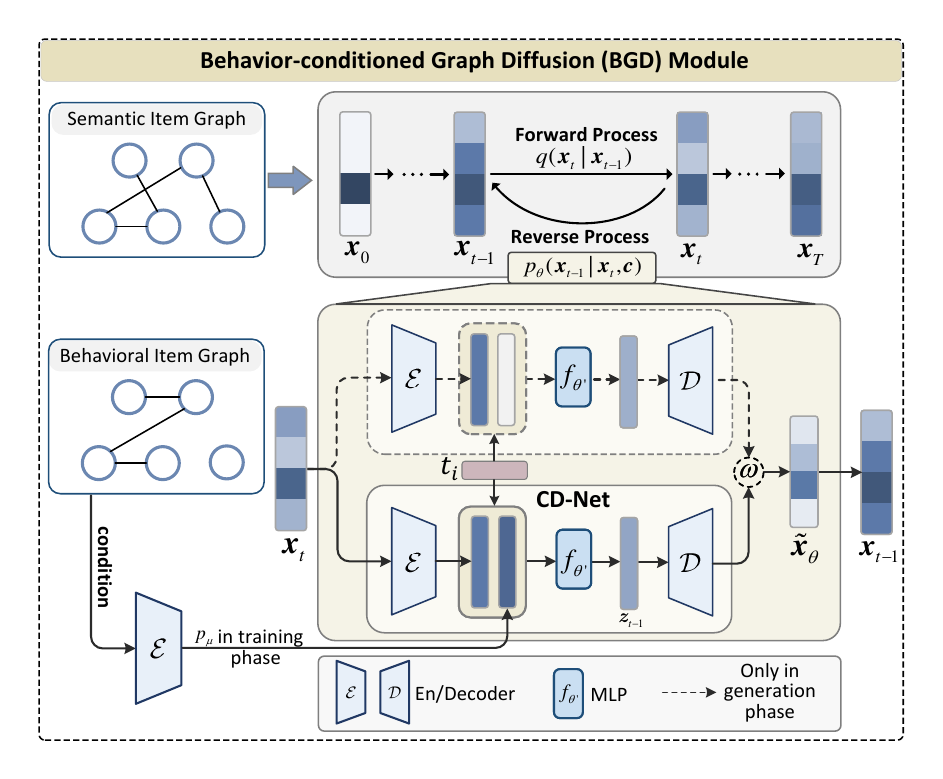}
    \textsf{\scriptsize (b) Structure of the BGD module.}
  \end{minipage}
  \caption{The structure overview of the proposed IGDMRec and the BGD module. \LL{IGDMRec first constructs semantic and behavioral item graphs, then employs the BGD module to perform behavior-conditioned diffusion and generate a diffusion-aware item graph. GCN-based propagation is subsequently conducted on the interaction graph and the two item graphs to obtain dual-view item representations, which are finally optimized via contrastive learning and BPR loss.}}
  \label{Fig:Diff}
\end{figure*}

\subsection{Notations}
Let $\mathcal{U}$ and $\mathcal{I}$ denote the sets of users and items, respectively. The user-item interaction matrix is $\boldsymbol{A}\in\mathbb{R}^{|\mathcal{U}|\times|\mathcal{I}|}$, where $|\mathcal{U}|$ and $|\mathcal{I}|$ denote the number of users and items, respectively, and $\boldsymbol{A}_{ui}=1$ suggests that user $u$ interacts with item $i$, otherwise $\boldsymbol{A}_{ui}=0$. Let the user-item graph be denoted as $\mathcal{G}=(\mathcal{V},\mathcal{O})$, where $|\mathcal{V}|=|\mathcal{U}|+|\mathcal{I}|$. Here, $\mathcal{V}$ represents the set of nodes, and $\mathcal{O}$ represents the set of edges. We consider two modalities $\mathcal{M}=\{v_m,t_m\}$ in this paper, where $v_m$ and $t_m$ represent the visual modality and the textual modality, respectively.

\section{Methodology}
In this section, we describe our proposed IGDMRec, which incorporates potential item relationships from the user-item interaction graph into the reconstruction of the semantic item graph through a conditional DM and leverages the diffusion-aware item graph to achieve representation augmentation for accurate recommendations. As shown in Fig.~\ref{Fig:Diff}(a), IGDMRec consists of four processes. Firstly, we construct the semantic item graph based on modality feature similarity and the behavioral item graph from higher-order item relationships within the interaction matrix. Then, a BGD module is proposed to generate a diffusion-aware item graph that integrates both semantic and behavioral information. As shown in Fig.~\ref{Fig:Diff}(b), we design a CD-Net to implement the denoising process with manageable complexity. Subsequently, we leverage Graph Convolutional Networks (GCNs) to extract user and item representations from the interaction graph and item graphs, and contrastive learning is introduced to enhance item representations. Finally, the recommendation task is optimized through the joint utilization of contrastive learning loss and Bayesian personalized ranking loss.

\subsection{Graph Construction}

\subsubsection{Semantic Item Graph}
We construct the semantic item graph for each modality $m\in\{v_m,t_m\}$ by evaluating the similarity between the raw features of the items. Specifically, the similarity score matrix $\tilde{\boldsymbol{S}}^m$ in modality $m$ is calculated by the cosine similarity function, with each element given by:
\begin{equation}
    \tilde{ S}_{ij}^{m}=\frac{(\boldsymbol{g}_i^m)^{\text{T}}\boldsymbol{g}_j^m}{\|\boldsymbol{g}_i^m\|\|\boldsymbol{g}_j^m\|},
\end{equation}
where $\boldsymbol{g}_i^m$ and $\boldsymbol{g}_j^m$ are the raw features of items $i$ and $j$ in modality $m$, respectively. Then, we employ the K-Nearest Neighbors (KNN) method~\cite{KNN} for each item to obtain the semantic item graph $\boldsymbol{S}^m$ in modality $m$, with each element given as:
\begin{equation}
\label{Eqn:KNN}
    S_{ij}^{m} =\left\{
    \begin{array}{ll}
     1, & \tilde{S}_{ij}^m\in \text{top-}k(\tilde{\boldsymbol{S}}_{:,j}^m), \\
     0, & \text{otherwise},
    \end{array}\right.
\end{equation}
where $\tilde{\boldsymbol{S}}_{:,j}^m$ denotes the $j$-th column of $\tilde{\boldsymbol{S}}^m$, $S_{ij}^{m}=1$ indicates a potential semantic relationship between item pair $i$ and $j$ in modality $m$, and $0$ otherwise. By aggregating the semantic item graphs of each modality, the adjacency matrix $\boldsymbol{S}$ of the final semantic item graph is given by:
\begin{equation}
\label{Eqn:item_graph_weight}
    \boldsymbol{S}=\sum_{m\in\mathcal{M}}\phi_m{\boldsymbol{S}}^m,
\end{equation}
where $\phi_m$ is the hyperparameter that controls the importance of modality $m$, subject to the constraint $\textstyle{\sum}_{m\in\mathcal{M}}\phi_m=1$.

\subsubsection{Behavioral Item Graph}
To capture item relationships driven by user preferences, we construct the behavioral item graph from the interaction matrix. Specifically, the relationship strength between an item pair $i$ and $j$ is measured by their co-occurrence frequency in user interactions, denoted as $\tilde{S}_{ij}^c$:
\begin{equation}
    \tilde{S}_{ij}^c=|\boldsymbol{U}_i\cap\boldsymbol{U}_j|,
\end{equation}
where $\boldsymbol{U}_i$ and $\boldsymbol{U}_j$ are the sets of users who have interacted with items $i$ and $j$, respectively. To mitigate the impact of noise in the interaction data, we prune low-valued elements while preserving the connectivity of the top-$k$ edges for each item. On this basis, the adjacency matrix $\boldsymbol{S}^c$ of the behavioral item graph can be obtained, with each element given as:
\begin{equation}
\label{Eqn:S^c}
    S_{ij}^c=\left\{
    \begin{array}{ll}
     \tilde{S}_{ij}^c, & \tilde{S}_{ij}^c\in \text{top-}k(\tilde{\boldsymbol{S}}_{:,j}^c)\ \& \ i\neq j\ \& \ \tilde{S}_{ij}^c>\varepsilon,\\
     1, & i=j,\\
     0, & \text{otherwise},
    \end{array}\right.
\end{equation}
where $\varepsilon$ denotes the pruning threshold.

\begin{algorithm}[t]
    \caption{BGD-Learning Phase (Single Batch)}
    \label{Al:learn}
    \begin{algorithmic}[1]
    \Require{Adjacency matrices $\boldsymbol{S}$, $\boldsymbol{S}^c$, and parameters $\theta$.}
        \State {Sample an item subset $\mathcal{I}_b \subset \mathcal{I}$.}
        \State {Get relation vectors: 
        $\boldsymbol{X}\!\!=\!\!\{\boldsymbol{S}_{:,i}| i\in\mathcal{I}_b\}$, 
        $\boldsymbol{C}\!\!=\!\!\{\boldsymbol{S}^c_{:,i}| i\in\mathcal{I}_b\}$.}
        \For{{$j = 1,\dots,|\mathcal{I}_b|$}} 
            \State {$\boldsymbol{x}_0 \gets \boldsymbol{S}_{:,j}$,\quad $\boldsymbol{c} \gets \boldsymbol{S}^c_{:,j}$}
            \State Sample $t \sim \mathrm{Uniform}(\{1,\dots,T\})$, $\boldsymbol{\epsilon}\sim\mathcal{N}(\boldsymbol{0},\boldsymbol{I})$;
            \State Compute $\boldsymbol{x}_t$ from $\boldsymbol{x}_0$, $t$, $\boldsymbol{\epsilon}$ by Eq.~\eqref{Eqn:x_t-x_0};
            \State Set $\boldsymbol{c} \leftarrow \boldsymbol{0}$ with probability $p_{\mu}$;
            \State Compute $\mathcal{L}_{t}^{\mathrm{DM}}$ by Eq.~\eqref{Eqn:lossdiff-1}--\eqref{Eqn:lossdiff-2};
            \State Update $\theta \leftarrow \theta - \eta\,\nabla_{\theta}\mathcal{L}_{t}^{\mathrm{DM}}$ \Comment{$\eta$ is the step size.}
        \EndFor
    \Ensure{Optimized $\theta$.}
    \end{algorithmic}
\end{algorithm}

\subsection{Diffusion-aware Graph Generation} 
\label{BGD}
In this section, we will detail the BGD module based on the conditional DM in Section \ref{Sec:CDM}. As illustrated in Fig.~\ref{Fig:Diff}(b), the behavioral item graph is incorporated as conditioning information into the denoising process of the semantic item graph, aiming to generate item relationships that integrate semantic with behavioral information. Specifically, during the learning phase of BGD, forward and reverse processes are employed to capture the item graph structure by optimizing model parameters. In the subsequent generation phase, the diffusion-aware relationship vectors are constructed based on the optimized model. 

\subsubsection{Learning Phase}
Given a relationship vector $\boldsymbol{S}_{:,j}$ for a specific item $j$ extracted from the semantic item adjacency matrix $\boldsymbol{S}$ in Eq.~\eqref{Eqn:item_graph_weight}, we set $\boldsymbol{x}_0=\boldsymbol{S}_{:,j}$ as the initial state to perform forward diffusion $\boldsymbol{x}_1\rightarrow\boldsymbol{x}_2\rightarrow\cdots\rightarrow\boldsymbol{x}_T$ for $T$ steps, as defined in Eq.~\eqref{eqn:Diff-Forward}. Following~\cite{DIFFREC}, a linear noise schedule for $1-\bar{\alpha}_t$ is employed to mitigate the excessive degradation of sparse information inherent in recommender systems. Specifically, for $t\in[1,T]$,
\begin{equation}
    1-\bar{\alpha}_t=s\cdot\left[\alpha_{\text{min}}+\frac{t-1}{T-1}(\alpha_{\text{max}}-\alpha_{\text{min}})\right], 
\end{equation}
where $s\in[0,1]$ is a hyperparameter that controls the noise scale, and $\alpha_{\text{min}}$ and $\alpha_{\text{max}}$ are the hyperparameters indicating the lower and upper bounds of the added noise, respectively.

In the reverse process outlined in Eq.~\eqref{Eqn:reverse}, the corresponding behavioral item relationship vector calculated by Eq.~\eqref{Eqn:S^c} is utilized as the conditioning information $\boldsymbol{c}=\boldsymbol{S}^c_{:,j}$ to guide the denoising of the semantic item relationship vector. Following \eqref{Eqn:diffL1} and \eqref{Eqn:diffLt}, the loss function for the learning phase in BGD is given as: for $\forall t\in[1,T]$,
\begin{equation}
    \label{Eqn:lossdiff-1}\mathcal{L}_{t}^{\text{DM}}=\mathbb{E}_{q(\boldsymbol{x}_t|\boldsymbol{x}_0)}\big[\hat{\alpha}_t\|\boldsymbol{x}_0-\hat{\boldsymbol{x}}_{\theta}(\boldsymbol{x}_t,\boldsymbol{c},t)\|_2^2\big],
\end{equation}
with
\begin{equation}
\label{Eqn:lossdiff-2}
        \hat{\alpha}_t=\left\{
    \begin{array}{ll}
     1,&t=1,\\
     \tfrac{1}{2}\left(\tfrac{\bar{\alpha}_{t-1}}{1-\bar{\alpha}_{t-1}}-\tfrac{\bar{\alpha}_t}{1-\bar{\alpha}_t}\right),&t\geq2,
    \end{array}\right.
\end{equation}
where the architecture of $\hat{\boldsymbol{x}}_{\theta}(\cdot)$ is detailed in Section \ref{Sec:CD-NET}.


\LL{In practice, we adopt a batch-wise item sampling strategy to improve training efficiency. Specifically, in each batch, we sample a subset of items $\mathcal{I}_b\subset\mathcal{I}$ and extract their corresponding relationship vectors from both the semantic graph $\boldsymbol{S}$ and the behavioral graph $\boldsymbol{S}^c$. This process enables BGD to learn an efficient denoising function at the vector level.} During optimization, we uniformly sample the time step $t$ to optimize the loss function. In addition, following the classifier-free guidance scheme~\cite{class-free}, the conditioning information $\boldsymbol{c}$ is randomly replaced with an empty token $\boldsymbol{0}$ with probability $p_{\mu}$, enabling the joint training of both conditional and unconditional DMs. See Algorithm \ref{Al:learn} for more details about the learning phase of BGD.

\begin{algorithm}[t]
    \caption{BGD-Generation Phase}
    \label{AL:gen}
    \begin{algorithmic}[1]
        \Require{$\boldsymbol{x}_T=\boldsymbol{S}_{:,j}$ and $\boldsymbol{c}=\boldsymbol{S}^c_{:,j}$ for item $j$, and optimized $\theta$.}
        \State Sample $\boldsymbol{\epsilon}\sim\mathcal{N}(\boldsymbol{0},\boldsymbol{I})$.
        \For{$t=T,\dots,1$}
        \State Get the estimation $\hat{\boldsymbol{x}}_{\theta}(\boldsymbol{x}_t,\boldsymbol{c},t)$ and $\hat{\boldsymbol{x}}_{\theta}(\boldsymbol{x}_t,\boldsymbol{0},t)$;
        \State Calculate $\tilde{\boldsymbol{x}}_{\theta}(\boldsymbol{x}_t,\boldsymbol{c},t)$ by Eq.~\eqref{Eqn:omega}; 
        \State Calculate $\boldsymbol{x}_{t-1}$ by Eq.~\eqref{Eqn:BGD-reverse};
        \EndFor
        \Ensure{Denoising relationship vector $\boldsymbol{x}_0$ for item $j$.}
    \end{algorithmic}
\end{algorithm}

\subsubsection{Generation Phase}
In the generation phase, we target to generate diffusion-aware relationship vectors for each item. Following classifier-free guidance~\cite{class-free}, we control the effect of the conditioning information $\boldsymbol{c}$ by modifying the prediction $\hat{\boldsymbol{x}}_{\theta}(\boldsymbol{x}_t,\boldsymbol{c},t)$ as:
\begin{equation}
\label{Eqn:omega}
    \tilde{\boldsymbol{x}}_{\theta}(\boldsymbol{x}_t,\boldsymbol{c},t) = (1+\omega) \hat{\boldsymbol{x}}_{\theta}(\boldsymbol{x}_t,\boldsymbol{c},t) -\omega\hat{\boldsymbol{x}}_{\theta}(\boldsymbol{x}_t,\boldsymbol{0},t),
\end{equation}
where $\omega$ is a hyperparameter that controls the strength of $\boldsymbol{c}$, attaining a trade-off between behavioral and semantic information in the item graph generation phase. 

Following~\cite{DDPM}, we ignore the variance $\Sigma_{\theta}(\boldsymbol{x}_t,\boldsymbol{c},t)$ in Eq.~\eqref{Eqn:reverse} and utilize $\boldsymbol{x}_{t-1}=\mu_{\theta}(\boldsymbol{x}_t,\boldsymbol{c},t)$ for deterministic inference. Thus, based on Eq.~\eqref{Eqn:mu} and Eq.~\eqref{Eqn:omega}, the one-step reverse process is given as:
\begin{equation}
\label{Eqn:BGD-reverse}
    \boldsymbol{x}_{t-1}=\frac{\sqrt{\alpha_t}(1-\bar{\alpha}_{t-1})}{1-\bar{\alpha}_t}\boldsymbol{x}_t+\frac{\sqrt{\bar{\alpha}_{t-1}}(1-\alpha_t)}{1-\bar{\alpha}_t}\tilde{\boldsymbol{x}}_{\theta}(\boldsymbol{x}_t,\boldsymbol{c},t).
\end{equation}
During the generation phase, for each item $j$, we set the initial state $\boldsymbol{x}_{T}=\boldsymbol{S}_{:,j}$ to perform reverse denoising $\boldsymbol{x}_{T}\rightarrow\boldsymbol{x}_{T-1}\rightarrow\cdots\rightarrow\boldsymbol{x}_0$ for $T$ steps, and finally obtain the denoising relationship vector $\boldsymbol{x}_0^j=\boldsymbol{x}_0$ for item $j$. Note that the omission of the forward process is intentional to prevent the corruption of semantic information.

Algorithm \ref{AL:gen} shows the details of BGD's generation phase. Subsequently, we use the KNN method to construct the adjacency matrix $\boldsymbol{S}^{d}$ of the diffusion-aware item graph, with each element given as:
\begin{equation}
\label{Eqn:KNN-2}
    S_{ij}^{d} =\left\{
    \begin{array}{ll}
     1, & (\boldsymbol{x}_0^j)_i\in \text{top-}k({\boldsymbol{x}}_{0}^j), \\
     0, & \text{otherwise},
    \end{array}\right.
\end{equation}
where $(\boldsymbol{x}_0^j)_i$ denotes the $i$-th element of the denoising relationship vector $\boldsymbol{x}_0^j$ for item $j$, which is obtained by Algorithm~\ref{AL:gen}.


\subsubsection{CD-Net}
\label{Sec:CD-NET}
The prediction $\hat{\boldsymbol{x}}_{\theta}(\boldsymbol{x}_t,\boldsymbol{c},t)$ in Eq.~\eqref{Eqn:lossdiff-1} is realized by the designed Conditional Denoising Network (CD-Net), which includes a codec system and a Multi-Layer Perceptron (MLP). As illustrated in Fig.~\ref{Fig:Diff}(b), an encoder $\mathcal{E}(\cdot)$ is employed to transform high-dimensional item-relationship vectors to a low-dimensional latent space, thus maintaining manageable complexity by controlling the latent dimension. Similar to~\cite{CF-Diff}, the encoder $\mathcal{E}(\cdot)$ is implemented as a linear transformation, i.e.,
\begin{equation}
    \boldsymbol{z}_t=\mathcal{E}(\boldsymbol{x}_t)=\boldsymbol{E}\boldsymbol{x}_t,
\end{equation}
where $\boldsymbol{E}\in\mathbb{R}^{k_d\times|\mathcal{I}|}$ represents the transformation matrix, and $k_d$ is the latent dimension. Similarly, the conditioning information in the latent space is represented as $\hat{\boldsymbol{c}}=\mathcal{E}(\boldsymbol{c})$. Then, the components $\boldsymbol{z}_t$, $\hat{\boldsymbol{c}}$, and the time step embedding $\boldsymbol{t}_i$ are concatenated and fed into the MLP $f_{{\theta'}}(\cdot)$ to predict $\boldsymbol{z}_{t-1}$. Here, $\boldsymbol{t}_i$ is generated from the scalar diffusion step $t$ by the sinusoidal embedding technique. Subsequently, the decoder $\mathcal{D}(\cdot)$ is designed to recover the prediction of $\boldsymbol{x}_0$:

\begin{equation}
    \hat{\boldsymbol{x}}_{\theta}(\boldsymbol{x}_t,\boldsymbol{c},t)=\mathcal{D}(\boldsymbol{z}_{t-1})=\boldsymbol{D}\boldsymbol{z}_{t-1},
\end{equation}
where $\boldsymbol{D}\in\mathbb{R}^{|\mathcal{I}|\times k_d}$ represents the transformation matrix.

\subsection{Contrastive Representation Augmentation }
In this section, GCNs are used to extract user and item representations from the interaction graph and item graphs. In addition, contrastive learning is employed to compare dual views of the final item representations, thus effectively utilizing potential relationships between items in the diffusion-aware item graph for representation augmentation.
\subsubsection{Information Propagation and Aggregation}
We construct a symmetric adjacency matrix $\boldsymbol{A}^{\rho}\in\mathbb{R}^{|\mathcal{V}|\times|{\mathcal{{V}}|}}$ by:
\begin{equation}
   \boldsymbol{A}^{\rho} =
\left(\begin{array}{ccc} 
 \boldsymbol{0}
&\boldsymbol{A}   \\
\boldsymbol{A}^{\text{T}} &\boldsymbol{0}
\end{array}\right),
\end{equation}
where $\boldsymbol{A}$ denotes the user-item interaction matrix. Then, $\boldsymbol{A}^{\rho}$ is normalized as $\hat{\boldsymbol{A}}^{\rho}=(\boldsymbol{D})^{-\frac{1}{2}}\boldsymbol{A}^{\rho}(\boldsymbol{D})^{-\frac{1}{2}}$, where $\boldsymbol{D}\in\mathbb{R}^{|\mathcal{V}|\times |\mathcal{V}|}$ is the diagonal degree matrix of $\boldsymbol{A}^{\rho}$ with $D_{ii}=\textstyle{\sum}_j{A}^{\rho}_{ij}$. Similarly, the semantic item graph $\boldsymbol{S}^m$ in Eq.~\eqref{Eqn:KNN} and the diffusion-aware item graph $\boldsymbol{S}^d$ in Eq.~\eqref{Eqn:KNN-2} are normalized as $\hat{\boldsymbol{S}}^m$ and $\hat{\boldsymbol{S}}^d$, respectively. Finally, the weighted semantic item graph is $\hat{\boldsymbol{S}}=\sum{m\in\mathcal{M}}\phi_m\hat{\boldsymbol{S}}^m$.

Then, we leverage LightGCN~\cite{LightGCN} for information propagation and aggregation on the three aforementioned graphs, i.e., $\smash{\hat{\boldsymbol{A}}^{\rho}}$, $\smash{\hat{\boldsymbol{S}}^d}$, and $\smash{\hat{\boldsymbol{S}}}$. Specifically, the feed-forward propagation over the normalized adjacency matrix $\hat{\boldsymbol{S}}$ of the semantic item graph  is defined as:
\begin{equation}
    \left(\boldsymbol{h}^{m}_{i}\right)^{l} = \sum_{j\in\mathcal{N}(i)}\hat{S}_{ij} \left(\boldsymbol{h}^{m}_{j}\right)^{l-1},
\end{equation}
where $\mathcal{N}(i)$ is the neighbor items of item $i$, $\left(\boldsymbol{h}^{m}_{i}\right)^{l}\in\mathbb{R}^d$ represents the $l$-th layer item semantic representation of item $i$, and $\left(\boldsymbol{h}^{m}_{i}\right)^{0}$ denotes the corresponding ID embedding vector of item $i$. Here, $d$ is the dimension of an item or user ID embedding. Let $L^{ii}$ be the number of semantic item graph convolutional layers. We select representation from the last layer as the item semantic representation:
\begin{equation}
    \boldsymbol{h}_i^{m}=\left(\boldsymbol{h}^{m}_{i}\right)^{L^{ii}}.
\end{equation}
Analogously, we apply $L^{ii}$ and $L^{ui}$ layers of graph convolution operations to $\smash{\hat{\boldsymbol{S}}^d}$ and $\smash{\hat{\boldsymbol{A}}^{\rho}}$, respectively. This results in the diffusion-aware item representation $\boldsymbol{h}_i^{d}\in\mathbb{R}^d$, the item interaction representation $\boldsymbol{h}_i^{ui}\in\mathbb{R}^d$, and the user interaction representation $\boldsymbol{h}_u^{ui}\in\mathbb{R}^d$. 

Finally, we use the user interaction representation as its final representation $\boldsymbol{h}_u$. For the item, we sum $\boldsymbol{h}_i^m$ and $\boldsymbol{h}_i^d$ with $\boldsymbol{h}_i^{ui}$, respectively, to obtain its final representation $\hat{\boldsymbol{h}}_i$ and its diffusion-aware final representation $\boldsymbol{h}_i$, i.e.,
\begin{equation}
    \boldsymbol{h}_u = \boldsymbol{h}_u^{ui}, \quad
    \hat{\boldsymbol{h}}_i = \boldsymbol{h}_i^{ui}+\boldsymbol{h}_i^m, \quad
    {\boldsymbol{h}}_i = \boldsymbol{h}_i^{ui}+\boldsymbol{h}_i^d.
\end{equation}

\subsubsection{Contrastive Learning for Representation Augmentation}
Treat the final representations of the items and their diffusion-aware final representations as dual views. To promote the exploration of item relationships, a self-supervised learning task based on contrastive learning has been devised with the goal of maximizing the agreement between positive pairs and minimizing the agreement between negative pairs. The contrastive learning loss is given as:
\begin{equation}
    \mathcal{L}_{\text{CL}}=\sum_{i\in \mathcal{I}}-\log\frac{\exp(s(\hat{\boldsymbol{h}}_i,\boldsymbol{h}_i)/\tau)}{\sum_{v\in \mathcal{I}}\exp(s(\hat{\boldsymbol{h}}_i,\boldsymbol{h}_v)/\tau)},
\end{equation}
where ($(\hat{\boldsymbol{h}}_i,\boldsymbol{h}_i)|i\in\mathcal{I}$) are positive pairs,  ($(\hat{\boldsymbol{h}}_i,\boldsymbol{h}_v)|i,v\in\mathcal{I},i\neq v$) are negative pairs, $s(\cdot)$ is the cosine similarity function, and $\tau$ is the temperature hyperparameter.

\subsection{Optimization}
Following the classical recommendation algorithms~\cite{freedom,LATTICE,BPR-MF}, we adopt the Bayesian Personalized Ranking (BPR) loss as follows:
\begin{equation}
\label{Eqn:bpr}
    \mathcal{L}_{\text{BPR}}=\sum_{(u,i,j)\in\mathcal{R}}(-\log\sigma(\boldsymbol{h}_u^{\text{T}}\boldsymbol{h}_i-\boldsymbol{h}_u^{\text{T}}\boldsymbol{h}_j)),
\end{equation}
where $\mathcal{R}$ is the training set, with each user-item triplet $(u,i,j)$ satisfying $\boldsymbol{A}_{ui}=1$ and $\boldsymbol{A}_{uj}=0$, and $\sigma(\cdot)$ denotes the sigmoid function. With these definitions, the integrative optimization loss for the recommendation task is:
\begin{equation}
\mathcal{L}_{\text{Rec}}=\mathcal{L}_{\text{BPR}}+\lambda_1\mathcal{L}_{\text{CL}}+\lambda_2\|\Theta\|_2^2,
\end{equation}
where $\|\Theta\|_2^2$ is the $L_2$ regularization term, and $\lambda_1$ and $\lambda_2$ are hyperparameters used to control the strength of the contrastive learning loss and the $L_2$ regularization term, respectively.

\section{Experiments}
\LL{In this section, we conduct extensive experiments to answer the following five key research questions (RQs):}
\begin{itemize}
    \item \textbf{RQ1}: How does IGDMRec perform compared with the state-of-the-art methods for recommendation?
    \item \textbf{RQ2}: How efficient is IGDMRec in terms of computational complexity and memory cost?
    \item \LL{\textbf{RQ3}: How robust is IGDMRec under incomplete or noisy multimodal conditions?}
    \item \textbf{RQ4}: How does each component in IGDMRec influence its recommendation accuracy?
    \item \textbf{RQ5}: How do key parameters influence the results?
\end{itemize}

\subsection{Experimental Settings}

\subsubsection{Datasets}
\LL{We conduct experiments on four widely used multimodal recommendation datasets, including three Amazon review datasets (Baby, Sports, Clothing) and the Allrecipes dataset. Dataset statistics are given in Table~\ref{tab:dataset}.}
\LL{
\begin{itemize}
    \item \textit{Amazon}: The Amazon review data provide visual and textual modalities for items, where each review rating is treated as a positive interaction. Following the common 5-core preprocessing, we use pre-trained CNNs to extract 4,096-dimensional visual features and sentence-transformers to obtain 384-dimensional textual embeddings, consistent with \cite{freedom}.
    \item \textit{Allrecipes}: The Allrecipes dataset contains 52,821 food recipes across 27 categories from a large food-oriented social platform. Each recipe is associated with an image and a list of ingredients, used as visual and textual features with dimensions 2,048 and 20, respectively. Following \cite{food}, 20 ingredients are sampled per recipe to construct the textual feature.
\end{itemize}
}


\subsubsection{Baselines}
To comprehensively demonstrate the superiority of our proposed IGDMRec, we compare it with several representative methods. These baselines are divided into two categories: General RSs, which rely solely on interactive data for recommendations, and MRSs, which utilize interactive data and multimodal features for recommendations.

\begin{itemize}
    \item {General RSs}: We have selected the two most representative recommendation methods for comparison: a conventional matrix factorization method (BPR-MF \cite{BPR-MF}) and a GCN-based method (LightGCN \cite{LightGCN}).
    \item {MRSs}: \LL{We have selected four types of MRSs: a) feature-based methods (VBPR \cite{VBPR} and MMGCN \cite{MMGCN}), b) SSL-based methods (SLMRec \cite{SLMRec}, and BM3 \cite{BM3}),} c) structure-based methods (LATTICE \cite{LATTICE} and FREEDOM \cite{freedom}), and d) diffusion-based methods (DiffMM \cite{DiffMM} and MCDRec \cite{MCDRec}).
\end{itemize}
Notably, IGDMRec focuses on enhancing structure-based MRSs by leveraging DMs. Thus, we emphasize comparisons with FREEDOM as a representative structure-based method and with recent diffusion-based MRSs like DiffMM and MCDRec to validate the advantages of our approach.

\subsubsection{Performance Metrics}
To evaluate the accuracy of our top-$K$ recommendation results, we utilize two commonly used metrics: Recall@K (R@K) and NDCG@K (N@K), where $K\in\{10,20\}$. For each user, all non-interacted items are ranked to compute these metrics.

\begin{table}
\centering 
  \caption{\LL{Statistics of the four multimodal datasets}}
  \label{tab:dataset}
  \begin{tabular}{lrrrr}
    \toprule
    Dataset& \#Users& \#Items& \#Interactions& Density\\
    \midrule
    Baby & 19,445 & 7,050 & 160,792&0.117\%\\
    Sports &35,598& 18,357&296,337&0.045\%\\
    Clothing & 39,387 & 23,033& 278,677&0.031\%\\
    \LL{Allrecipes} & \LL{19{,}805} & \LL{10{,}067} & \LL{58{,}922} & \LL{0.030\%} \\
  \bottomrule
\end{tabular}
\end{table}

\subsubsection{Implementation Details}
Following the existing work \cite{freedom}, the embedding size $d$ for both users and items is set to 64. In addition, the embedding parameters are initialized using the Xavier method \cite{Xavier}, and all models are optimized with the Adam optimizer at a learning rate of 0.001. We use either the original implementations of the methods or the implementations in MMRec \cite{MMRec} with default parameters. All methods are implemented using PyTorch 2.3.0 and Python 3.10.14, with experiments performed on an NVIDIA RTX 4090D GPU card with 24 GB of memory. An early stopping strategy is employed with a patience of 20 epochs, while the total number of training epochs is set to 1000. The stopping criterion is based on the R@20 metric. 

\subsubsection{Hyperparameters Settings}
We perform a comprehensive grid search to select the optimal hyperparameters. Specifically, the number of GCN layers for the user-item interaction graph and the item graph are set to $L_{ui}=2$ and $L_{ii}=1$, respectively. The hyperparameter $\lambda_2$ is empirically fixed at $1e^{-7}$, the visual feature ratio $\phi_{v_m}$ is set to 0.1, the value of $k$ for top-$k$ is fixed at 10, and the pruning threshold $\varepsilon$ is set to 2. For contrastive learning, the temperature parameter $\tau$ and the weight $\lambda_1$ are selected from $\{0.1,0.2,0.5,1\}$ and $\{1e^{-1},1e^{-2},1e^{-3}\}$, respectively. Regarding the conditional DM, the time step embedding size is fixed at $10$, while the latent dimension $k_d$ is chosen from $\{1000,2000,3000,4000\}$. The diffusion step $T$ are tuned within $\{2,5,10\}$, and the lower bounds of the added noise $\alpha_{min}$ is set empirically at $1e^{-4}$. The noise scale $s$ and the upper bounds of the added noise $\alpha_{max}$ are searched in $\{1e^{-2},2e^{-3}\}$ and $\{2e^{-2},5e^{-2}\}$, respectively. In addition, the probability $p_{\mu}$ is set to 0.1 following \cite{Genearate-what}, and the hyperparameter $\omega$ used to control the strength of the conditioning information is tuned in $\{0,2,4,6,8\}$.

\begin{table*}[!t]
  \caption{Overall performance comparison between the baselines and IGDMRec. The best and second-best performances are highlighted in bold and underlined, respectively. We conduct experiments across 5 different seeds and state the improvements over FREEDOM~\cite{freedom} are statistically significant at the level of $p<0.05$ with a paired $t$-test.}
  \label{tab:comparison}
  \begin{threeparttable}
  \setlength{\tabcolsep}{5pt}
  \resizebox{0.98\textwidth}{!}{
  \begin{tabular}{cc|ccccccccccccc}
    \toprule
    
    Dataset& Metric & BPR\text{-}MF & LightGCN & VBPR & MMGCN & SLMRec & BM3 & LATTICE & FREEDOM & DiffMM & MCDRec & IGDMRec & \textit{vs.FREEDOM}$\uparrow$ & \textit{vs.Best}$\uparrow$ \\
    \midrule[\heavyrulewidth]

    \multirow{4}{*}{Baby}
    & R@10 & 0.0357 & 0.0479 & 0.0422 & 0.0393 & 0.0549 & 0.0543 & 0.0544 & 0.0626 & 0.0604 & \underline{0.0644*} & \textbf{0.0675} & 7.83\% & 4.81\% \\
    & R@20 & 0.0575 & 0.0754 & 0.0664 & 0.0623 & 0.0838 & 0.0870 & 0.0864 & 0.0986 & 0.0942 & \underline{0.1013*} & \textbf{0.1055} & 7.00\% & 4.15\% \\
    & N@10 & 0.0192 & 0.0257 & 0.0223 & 0.0207 & 0.0295 & 0.0287 & 0.0288 & 0.0327 & 0.0319 & \underline{0.0343*} & \textbf{0.0366} & 11.93\% & 6.71\% \\
    & N@20 & 0.0249 & 0.0328 & 0.0285 & 0.0266 & 0.0370 & 0.0371 & 0.0366 & 0.0420 & 0.0406 & \underline{0.0438*} & \textbf{0.0464} & 10.48\% & 5.94\% \\
    \midrule

    \multirow{4}{*}{Sports}
    & R@10 & 0.0432 & 0.0569 & 0.0560 & 0.0369 & 0.0676 & 0.0646 & 0.0620 & 0.0724 & 0.0696 & \underline{0.0737*} & \textbf{0.0783} & 8.15\% & 6.24\% \\
    & R@20 & 0.0653 & 0.0864 & 0.0856 & 0.0602 & 0.1017 & 0.0977 & 0.0956 & 0.1097 & 0.1039 & \underline{0.1100*} & \textbf{0.1172} & 6.84\% & 6.55\% \\
    & N@10 & 0.0241 & 0.0311 & 0.0307 & 0.0186 & 0.0374 & 0.0353 & 0.0340 & 0.0390 & 0.0377 & \underline{0.0392*} & \textbf{0.0426} & 9.23\% & 8.67\% \\
    & N@20 & 0.0298 & 0.0387 & 0.0383 & 0.0246 & 0.0462 & 0.0438 & 0.0427 & 0.0486 & 0.0462 & \underline{0.0488*} & \textbf{0.0526} & 8.23\% & 7.79\% \\
    \midrule

    \multirow{4}{*}{Clothing}
    & R@10 & 0.0206 & 0.0361 & 0.0281 & 0.0211 & 0.0460 & 0.0415 & 0.0492* & \underline{0.0625} & 0.0567 & – & \textbf{0.0646} & 3.36\% & 3.36\% \\
    & R@20 & 0.0303 & 0.0544 & 0.0411 & 0.0345 & 0.0699 & 0.0620 & 0.0733* & \underline{0.0940} & 0.0848 & – & \textbf{0.0951} & 1.17\% & 1.17\% \\
    & N@10 & 0.0114 & 0.0197 & 0.0157 & 0.0108 & 0.0248 & 0.0226 & 0.0268* & \underline{0.0341} & 0.0302 & – & \textbf{0.0351} & 2.93\% & 2.93\% \\
    & N@20 & 0.0138 & 0.0243 & 0.0190 & 0.0142 & 0.0309 & 0.0278 & 0.0330* & \underline{0.0421} & 0.0372 & – & \textbf{0.0428} & 1.66\% & 1.66\% \\
    \midrule

    \multirow{4}{*}{\LL{Allrecipes}}
    & \LL{R@10} & \LL{0.0142} & \LL{0.0165} & \LL{0.0065} & \LL{0.0171} & \LL{0.0187} & \LL{\underline{0.0234}} & \LL{0.0151} & \LL{0.0139} & 0.0215 & \LL{–} & \LL{\textbf{0.0256}} & \LL{84.17\%} & \LL{9.40\%} \\
    & \LL{R@20} & \LL{0.0240} & \LL{0.0283} & \LL{0.0118} & \LL{0.0329} & \LL{0.0265} & \LL{\underline{0.0387}} & \LL{0.0262} & \LL{0.0202} & \LL{0.0298} & \LL{–} & \LL{\textbf{0.0431}} & \LL{113.37\%} & \LL{11.37\%} \\
    & \LL{N@10} & \LL{0.0067} & \LL{0.0085} & \LL{0.0033} & \LL{0.0069} & \LL{0.0093} & \LL{\underline{0.0111}} & \LL{0.0074} & \LL{0.0070} & \LL{0.0109} & \LL{–} & \LL{\textbf{0.0118}} & \LL{68.57\%} & \LL{6.31\%} \\
    & \LL{N@20} & \LL{0.0091} & \LL{0.0114} & \LL{0.0046} & \LL{0.0107} & \LL{0.0112} & \LL{\underline{0.0150}} & \LL{0.0102} & \LL{0.0086} & \LL{0.0131} & \LL{–} & \LL{\textbf{0.0161}} & \LL{87.21\%} & \LL{7.33\%} \\
    \bottomrule
  \end{tabular}
}
  \begin{tablenotes}
    \footnotesize
    \item * denotes results are copied from its original paper or FREEDOM. ‘-’ indicates that the original paper did not provide results and code implementation. 
  \end{tablenotes}
  \end{threeparttable}
\end{table*}

\subsection{Performance Comparison (RQ1\&RQ2)}
\subsubsection{Effectiveness (RQ1)} 
\label{Sec:pc}
To evaluate the effectiveness of IGDMRec, \LL{we conducted experiments on four datasets}, and the results presented in Table~\ref{tab:comparison} yield several key observations: 

\LL{\textbf{Consistent superiority across datasets}: IGDMRec exhibits consistent superiority over both general RSs and MRSs baselines across three Amazon categories and the non-Amazon Allrecipes dataset, confirming its ability to capture user preferences via explicit item graph denoising and dual-view contrastive learning. An interesting observation is that on Allrecipes, the SSL-based method BM3~\cite{BM3} outperforms FREEDOM~\cite{freedom}, which performs well on Amazon datasets but lacks an item graph denoising mechanism. This indicates that the semantic item graph constructed in the Allrecipes dataset is of lower quality due to the more pronounced modality noise, under which the inherent noise robustness of SSL-based methods exhibits a clear advantage. In contrast, IGDMRec performs best across all datasets, indicating that it not only inherits the noise-robustness of SSL-based methods via contrastive representation augmentation, but also benefits from explicitly mining latent item structures as in structure-based MRSs.}

\LL{\textbf{Benefits over structure-based MRSs (denoising effect)}: Structure-based MRSs (LATTICE~\cite{LATTICE} and FREEDOM~\cite{freedom}) generally outperform feature-based MRSs on three Amazon datasets, highlighting the value of explicitly modeling latent item structures. Compared with FREEDOM, which does not perform item graph denoising, IGDMRec consistently achieves stable gains. An exception occurs on the Clothing dataset, where the average gain is only 2.28\%, substantially smaller than on Baby and Sports. This likely reflects weaker modality noise and a more semantics-driven preference pattern in the Clothing dataset. In contrast, IGDMRec delivers a striking average improvement of 88.33\% over FREEDOM on Allrecipes, indicating that the semantic item graph in this dataset is highly noisy, while the behavior-conditioned denoising in IGDMRec effectively suppresses such noise and uncovers more reliable latent item structures.}

\LL{\textbf{Advantages over diffusion-based MRSs (latent structure mining)}: Compared with recent diffusion-based MRSs, IGDMRec yields consistent improvements. DiffMM~\cite{DiffMM} performs diffusion-based denoising on the user–item interaction graph with multimodal features, while MCDRec~\cite{MCDRec} incorporates multimodal signals into item embeddings via a diffusion process. In contrast, IGDMRec constitutes a principled structural enhancement over structure-based MRSs: it explicitly performs diffusion-based denoising on the semantic item graph and leverages the denoised graph into dual-view contrastive learning. The superiority of IGDMRec demonstrates the effectiveness of uncovering high-quality latent item structures for improving multimodal recommendation performance.}

\begin{table*}
\centering
\caption{Comparison of IGDMRec against state-of-the-art baselines on model efficiency}
\label{tab:eff}
\begin{threeparttable}
\begin{tabular}{ccccccccc}
\toprule
\multirow{2}{*}{Dataset} & \multirow{2}{*}{Metric} 
& \multicolumn{2}{c}{General RSs} 
& \multicolumn{2}{c}{Structure-based MRSs} 
& \multicolumn{3}{c}{Diffusion-based MRSs} \\
\cmidrule(lr){3-4} \cmidrule(lr){5-6} \cmidrule(lr){7-9} 
& & BPR-MF & LightGCN & LATTICE & FREEDOM & DiffMM & IGDMRec \\
\midrule[\heavyrulewidth]

\multirow{2}{*}{Baby} 
& Memory (GB) & 1.59 & 1.69 & 4.63 & 2.13 & 1.90 & 2.63 \\
& Time (s/epoch) & 1.38 & 1.32 & 1.66 & 1.58 & 3.83 & 2.47 \\
\midrule

\multirow{2}{*}{Sports} 
& Memory (GB) & 2.00 & 2.24 & 19.93 & 3.34 & 3.58 & 11.95 \\
& Time (s/epoch) & 2.32 & 2.44 & 9.48 & 3.52 & 12.89 & 8.28 \\
\midrule

\multirow{2}{*}{Clothing} 
& Memory (GB) & 2.16 & 2.43 & 28.22 & 4.15 & 3.75 & 17.63 \\
& Time (s/epoch) & 2.59 & 2.66 & -- & 3.86 & 14.97 & 13.62 \\
\bottomrule
\end{tabular}
\end{threeparttable}
\end{table*}

\begin{table}[!t]
  \caption{\LL{Performance and efficiency comparison between IGDMRec and IGDMRec*}}
  \label{tab:light}
  \centering
  \begin{threeparttable}
  \resizebox{0.49\textwidth}{!}{
  \begin{tabular}{c|c|lll}
    \toprule
    {Dataset} & {Method} & {R@10$\uparrow$} & {N@10$\uparrow$} & {Time (s/epoch)$\downarrow$} \\
    \midrule
    \multirow{2}{*}{{Sports}} 
        & IGDMRec & 0.0783 & 0.0426 & \ \quad8.28 \\
        & \cellcolor{gray!20}IGDMRec* & \cellcolor{gray!20}$0.0770^{\textcolor{darkgreen}{-1.7\%}}$ & \cellcolor{gray!20}$0.0420^{\textcolor{darkgreen}{-1.4\%}}$ & \cellcolor{gray!20}$\ \quad4.86^{\textcolor{blue}{-41.3\%}}$ \\
    \midrule
    \multirow{2}{*}{{Clothing}} 
        & IGDMRec & 0.0646 & 0.0351 & \quad13.62 \\
        & \cellcolor{gray!20}IGDMRec* & \cellcolor{gray!20}$0.0637^{\textcolor{darkgreen}{-1.4\%}}$ & \cellcolor{gray!20}$0.0344^{\textcolor{darkgreen}{-2.0\%}}$ &\cellcolor{gray!20} $\quad5.63^{\textcolor{blue}{-58.7\%}}$ \\
    \bottomrule
  \end{tabular}
  }
\end{threeparttable}
\end{table}

\subsubsection{Efficiency (RQ2)}


\LL{We report in Table~\ref{tab:eff} the comparison between IGDMRec and competitive baselines on memory consumption and running time across the three Amazon datasets. In general, MRSs incur higher memory and time costs than general RSs because they need to process and integrate multimodal information. Compared with existing structure-based and diffusion-based MRSs, IGDMRec maintains a manageable complexity, primarily due to the lightweight encoder–decoder design of the CD-Net. Instead of operating directly on the high-dimensional item relationship vector $\boldsymbol{x}_t \in \mathbb{R}^{|\mathcal{I}|\times 1}$, CD-Net projects it into a low-dimensional latent space $\boldsymbol{z}_t \in \mathbb{R}^{k_d\times 1}$ with $|\mathcal{I}| \gg k_d$, which effectively reduces both computation and memory usage.}

\begin{itemize}[leftmargin=*]
    \item \LL{\textit{Space complexity}: As can be seen in Table~\ref{tab:eff}, IGDMRec consumes less memory than LATTICE~\cite{LATTICE}, which also updates the item graph dynamically during training. Although FREEDOM~\cite{freedom} achieves lower memory usage by freezing the item graph, the performance gains delivered by IGDMRec justify its modest memory increase. The memory efficiency of CD-Net is empirically supported by Table~\ref{tab:ablation}: the \textit{w/o ED} variant (which removes the encoder–decoder) suffers from out-of-memory (OOM) issues on the Sports and Clothing datasets, while IGDMRec with CD-Net successfully avoids such failures, highlighting its memory efficiency.}
    \item \LL{\textit{Time complexity}: IGDMRec exhibits shorter running time than the recent diffusion-based method DiffMM~\cite{DiffMM}. Despite integrating a highly complex BGD module, its running time remains effectively controlled. The time complexity of each denoising step in CD-Net is $O(|\mathcal{I}|k_d + k_d^2)$, compared with $O(|\mathcal{I}|^2)$ for directly operating on the full item graph. Given that $|\mathcal{I}|$ in real-world datasets typically reaches hundreds of thousands, while $k_d$ is only on the order of thousands, this design substantially reduces the computational cost. Furthermore, adopting a smaller diffusion step size $T$ further ensures training efficiency.}
\end{itemize}

\LL{To further reduce time consumption, we additionally evaluate a lightweight variant IGDMRec* that updates the BGD module and the item graph once every 5 epochs instead of at every epoch. As shown in Table~\ref{tab:light}, this strategy significantly reduces the average running time per epoch (by 41.3\% on Sports and 58.7\% on Clothing) while causing only marginal performance drops (within 2\% on both R@10 and N@10). This demonstrates that diffusion-based denoising does not need to be executed in every epoch to remain effective, and can instead be spread over training to substantially enhance computational efficiency, thereby improving the practical applicability of IGDMRec in real-world scenarios.}

\vspace{-0.5cm}

\LL{\subsection{Robustness Analysis (RQ3)}}
\LL{To  validate the denoising capability of the BGD module, we conduct robustness experiments from two perspectives: 
1) noisy-modality scenarios, where Gaussian perturbations are injected into modality features, and 2) incomplete-modality scenarios, where partial modality features are missing. These experiments aim to evaluate whether IGDMRec can effectively preserve stable recommendation performance by leveraging its denoising ability when multimodal inputs are corrupted.}

\subsubsection{\LL{Robustness under noisy modalities}}
\LL{To assess the robustness of IGDMRec under noisy conditions, we follow the noise-injection protocol in~\cite{Mirror_gra} and add Gaussian noise $\tilde{\epsilon}\sim\mathcal{N}(0,10^{-4})$ to the multimodal features. We compare IGDMRec with FREEDOM~\cite{freedom}, a representative structure-based MRS that also constructs item graphs but lacks an explicit denoising mechanism, on three Amazon datasets. We then measure the average degradation of R@20 and N@20 relative to the noise-free setting (denoted as Avg.~$\Delta$). As shown in Table~\ref{tab:robust}, IGDMRec consistently outperforms FREEDOM, with notably smaller performance degradation across all Amazon datasets (Avg.~$\Delta$ reduced from 18.45\% to 13.82\% on {Baby}, from 20.31\% to 10.59\% on {Sports}, and from 45.28\% to 31.12\% on {Clothing}), demonstrating remarkable robustness to noisy multimodal features. This robustness benefits from the diffusion-based denoising mechanism, which explicitly mitigates stochastic perturbations in the semantic item graph and iteratively refines reliable item relationships under behavioral guidance, thereby ensuring stable recommendation performance under high-noise conditions.}

\subsubsection{\LL{Robustness under incomplete modalities}} 
\begin{figure}[tbp]
  \centering
  \includegraphics[width=\linewidth]{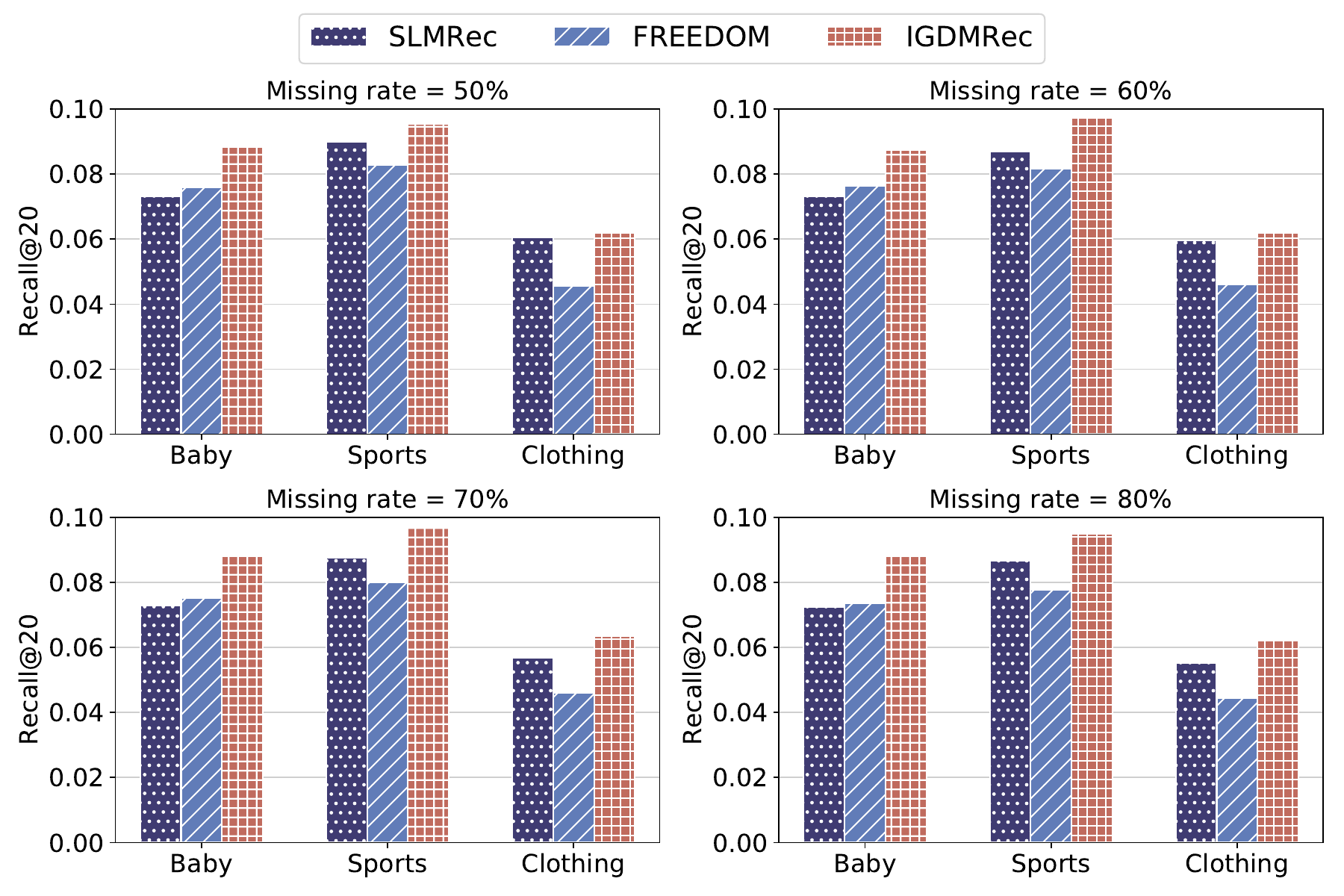}
  \caption{\LL{Performance about the comparison with SLMRec and FREEDOM with different missing rates for multimedia recommendation regarding Recall@20 of the IGDMRec on the Baby, Sports, and Clothing datasets.}}
  \label{Fig:missing}
\end{figure}

\LL{Following the setting in~\cite{Missing_rate}, we simulate incomplete-modality scenarios by randomly masking a certain proportion of visual and textual features, with missing rates varying from 50\% to 80\%. We compare IGDMRec with two representative baselines: the structure-based method FREEDOM~\cite{freedom} and the SSL-based method SLMRec~\cite{SLMRec}, and report Recall@20 results on three Amazon datasets. As shown in Fig.~\ref{Fig:missing}, IGDMRec achieves the best Recall@20 scores across all missing-rate settings (50\%, 60\%, 70\%, and 80\%) on all datasets. This indicates that the proposed diffusion-based denoising process can effectively reconstruct reliable semantic relationships even when a large portion of modality features is missing, thereby demonstrating strong robustness in incomplete-modality scenarios. This can be attributed to IGDMRec’s ability to effectively integrate behavioral item relationships to guide the diffusion process of the semantic item graph, resulting in stable and accurate graph reconstruction.}

\begin{table}[!t]
  \caption{\LL{Performance robustness under Gaussian noise injection ($\tilde{\epsilon}\sim\mathcal{N}(0,10^{-4})$)}}
  \label{tab:robust}
  \centering
  \begin{threeparttable}
  \begin{tabular}{c|c|ccc}
    \toprule
    {Dataset} & {Method} & {R@20$\uparrow$} & {N@20$\uparrow$} & Avg. $\Delta$ $\downarrow$ \\
    \midrule
    \multirow{2}{*}{{Baby}} 
      & FREEDOM  & 0.0803 & 0.0343 &  18.45\%  \\
      & IGDMRec  & \textbf{0.0909} & \textbf{0.0400} &  \textbf{13.82\%} \\
    \midrule
    \multirow{2}{*}{{Sports}} 
      & FREEDOM  & 0.0866 & 0.0391 &  20.31\% \\
      & IGDMRec  & \textbf{0.1042} & \textbf{0.0473} &  \textbf{10.59\%} \\
    \midrule
    \multirow{2}{*}{{Clothing}} 
      & FREEDOM  & 0.0513 & 0.0231 &   45.28\% \\
      & IGDMRec  & \textbf{0.0645} & \textbf{0.0297} &   \textbf{31.12\%} \\
    \bottomrule
  \end{tabular}
\end{threeparttable}
\end{table}

\subsection{Ablation Study (RQ4)}
\label{Sec:Ablation}

\begin{table}[t]
\centering
  \caption{Ablation study on key components of IGDMRec}
  \label{tab:ablation}
  \begin{threeparttable}
      \begin{tabular}{c|c|cc|cc}
        \toprule
        Dataset & Variant & R@10 & R@20 & N@10 & N@20 \\
        \midrule[\heavyrulewidth]
        \multirow{4}{*}{Baby} 
        & \emph{w/o CI}  & 0.0625 & 0.0983 & 0.0340 & 0.0432 \\
        & \emph{w/o CL} & 0.0636 & 0.0990 & 0.0341 & 0.0431 \\
        & \emph{w/o ED}  & 0.0657 & 0.1022 & 0.0355 & 0.0449 \\
        & IGDMRec  &\textbf{0.0675} & \textbf{0.1055} & \textbf{0.0366} & \textbf{0.0464}  \\
        \midrule
        \multirow{4}{*}{Sports} & \emph{w/o CI} & 0.0719 & 0.1090 & 0.0388 & 0.0484 \\
        & \emph{w/o CL}  & 0.0738 & 0.1113 & 0.0396 & 0.0493 \\
        & \emph{w/o ED}  & OOM & OOM & OOM & OOM  \\
        & IGDMRec  &  \textbf{0.0783} & \textbf{0.1172} & \textbf{0.0426} & \textbf{0.0526}  \\
        \midrule
        \multirow{4}{*}{Clothing} & \emph{w/o CI} & 0.0592 & 0.0867 & 0.0317 & 0.0387 \\
        & \emph{w/o CL}  & 0.0601 & 0.0905 & 0.0324 & 0.0401 \\
        & \emph{w/o ED}  & OOM & OOM & OOM & OOM \\
        & IGDMRec  &  \textbf{0.0646} & \textbf{0.0951} & \textbf{0.0351} & \textbf{0.0428}  \\
        \midrule
        \multirow{4}{*}{\LL{Allrecipes}} & \emph{\LL{w/o CI}} & \LL{0.0243} & \LL{0.0331} & \LL{0.0111} & \LL{0.0133} \\
        & \emph{\LL{w/o CL}}  & \LL{0.0254} & \LL{0.0390} & \LL{0.0117} & \LL{0.0148} \\
        & \emph{\LL{w/o ED}}  & \LL{0.0253} & \LL{0.0358} & \LL{0.0116} & \LL{0.0144} \\
        & \LL{IGDMRec}  &  \LL{\textbf{0.0256}} & \LL{\textbf{0.0431}} & \LL{\textbf{0.0118}} & \LL{\textbf{0.0161}}  \\
        \bottomrule
      \end{tabular}  
  \end{threeparttable}       
\end{table}

To evaluate the effectiveness of the different components in {IGDMRec}, we conduct an ablation study by individually removing three key components of {IGDMRec} as follows:
\begin{itemize}
    \item \emph{w/o CI}: This variant removes the conditioning information in the BGD module by setting $\omega=-1$ and $p_{\mu}=1$.
    \item \emph{w/o CL}: This variant removes the contrastive learning loss in the recommendation task.
    \item \emph{w/o ED}: This variant removes the encoder and decoder components in the CD-Net.
\end{itemize}

The performance comparison of IGDMRec and its variants is presented in Table \ref{tab:ablation}, we observed that: 1) The \emph{w/o CI} variant shows the largest performance degradation across all cases, highlighting the critical role of behavioral conditioning information in mining item relationships and the effectiveness of the BGD module in fusing semantic and behavioral information. 2) The \emph{w/o CL} variant also shows performance degradation across all cases, emphasizing the key role of contrastive learning in enabling representation augmentation. 3) \LL{The \emph{w/o ED} variant shows performance degradation on the Baby and Allrecipes datasets and encounters out-of-memory (OOM) on the Sports and Clothing datasets. This suggests that the codec system is crucial for capturing potential item relationships and controlling the memory usage of the CD-Net.}

\subsection{In-Depth Model Analysis (RQ5)}

\subsubsection{Effect of $\omega$ in the Denoising Process}
\begin{figure}[!tbp]
  \centering
  \includegraphics[width=\linewidth]{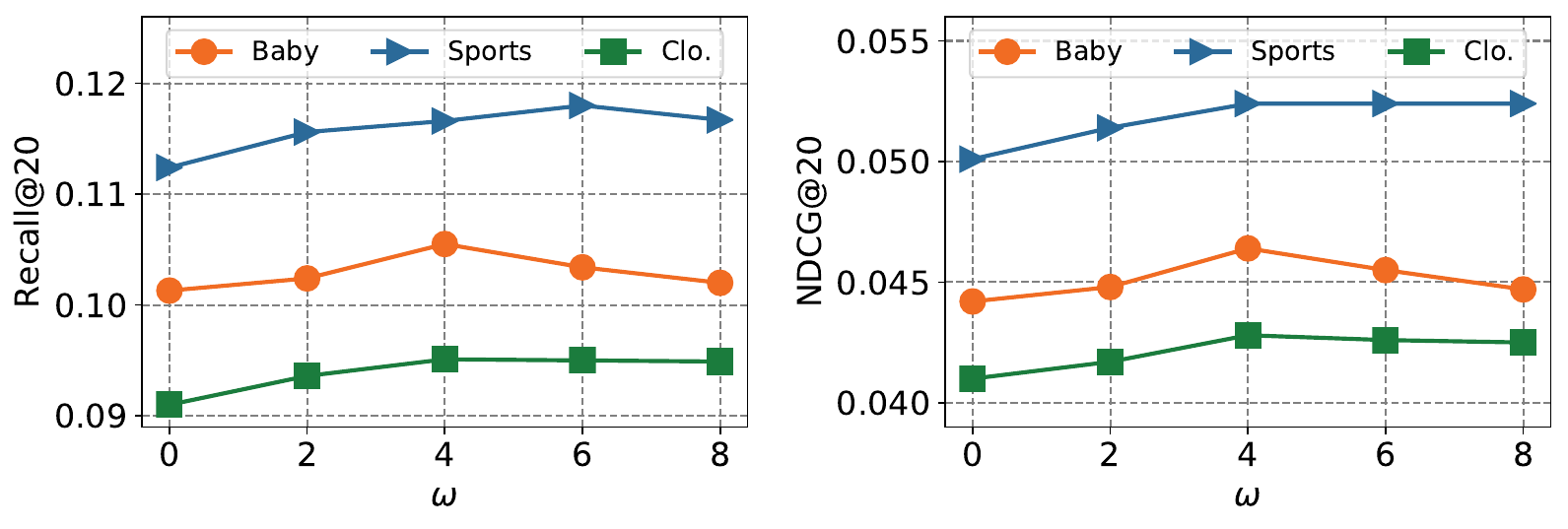}
  \caption{Performance comparison w.r.t different $\omega$.}
  \label{Fig:omega}
\end{figure}

The hyperparameter $\omega$ is introduced to control the strength of the conditioning information. To enable controlled guidance in IGDMRec, the classifier-free guidance scheme is employed by jointly training both a conditional and an unconditional DM. 
As shown in Fig.~\ref{Fig:omega}, we conduct experiments by adjusting the value of $\omega$ in Eq.~\eqref{Eqn:omega}, yielding the following observations: The poor performance observed across all three datasets when $\omega = 0$, where the predictions rely solely on the conditional DM (see Eq.~\eqref{Eqn:omega}). This {highlights the importance of the classifier-free guidance scheme in improving the quality of the item graph}. In addition, increasing the value of $\omega$ initially improves recommendation accuracy, aligning with the intuition that stronger conditioning facilitates the fusion of semantic and behavioral information. However, further increases in $\omega$ may result in a decline in performance, indicating that overemphasizing conditioning information can compromise the quality of item graph generation, thereby negatively impacting recommendation performance. \LL{Although the Allrecipes dataset is not depicted in Fig.~\ref{Fig:omega}, it exhibits a consistent trend and achieves its best performance at $\omega = 4$.}

\begin{figure}[!tbp]
  \centering
  \includegraphics[width=\linewidth]{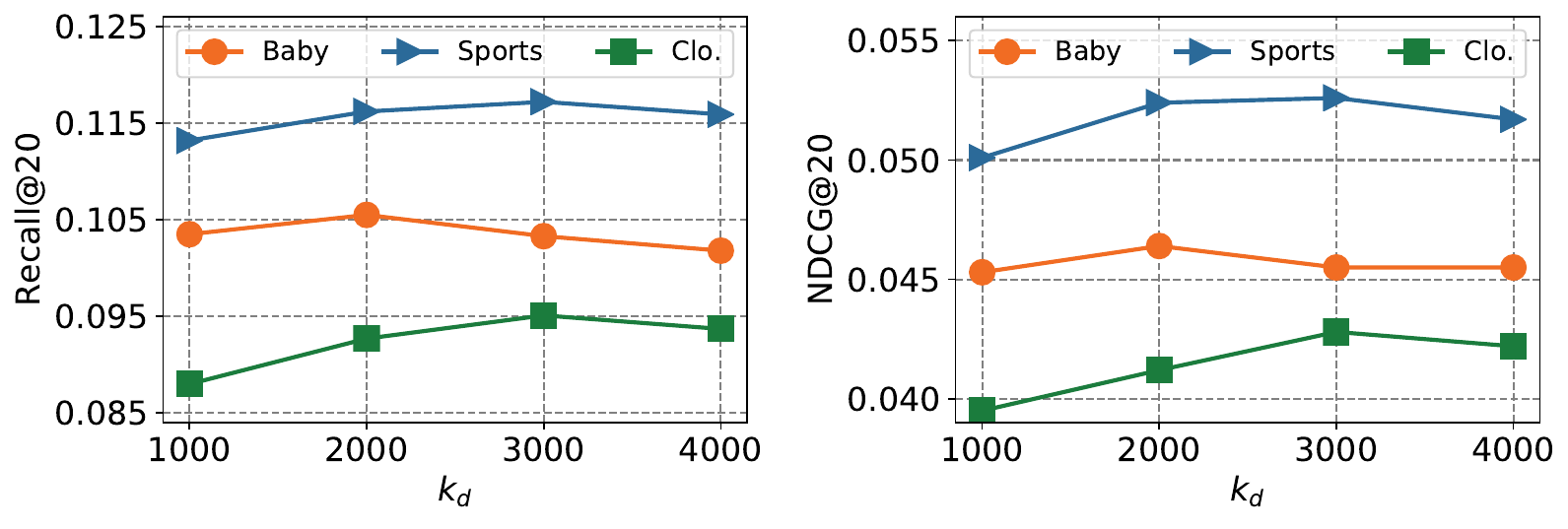}
  \caption{Performance comparison w.r.t different $k_d$.}
  \label{Fig:K}
\end{figure}

\begin{figure}[!tbp]
    \centering
    \begin{minipage}[t]{0.47\textwidth}
        \centering
        \includegraphics[width=\textwidth]{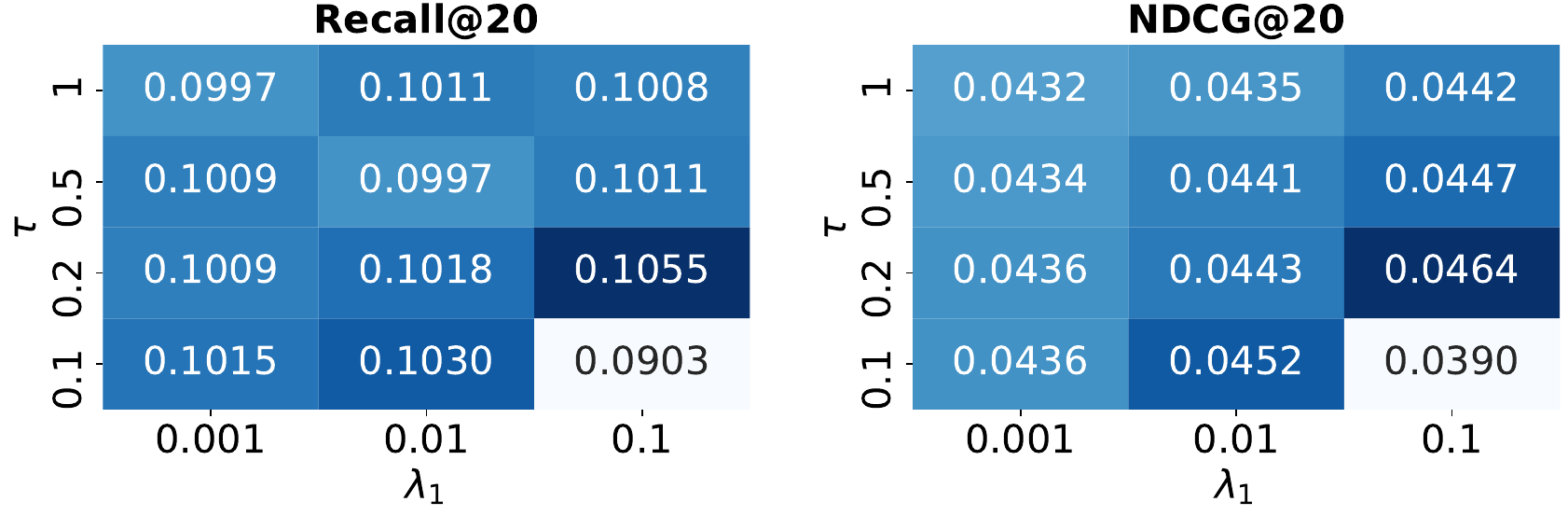}
        \textsf{\scriptsize (a) Baby}
    \end{minipage}

    \vspace{0.2cm}
    \begin{minipage}[t]{0.47\textwidth}
        \centering
        \includegraphics[width=\textwidth]{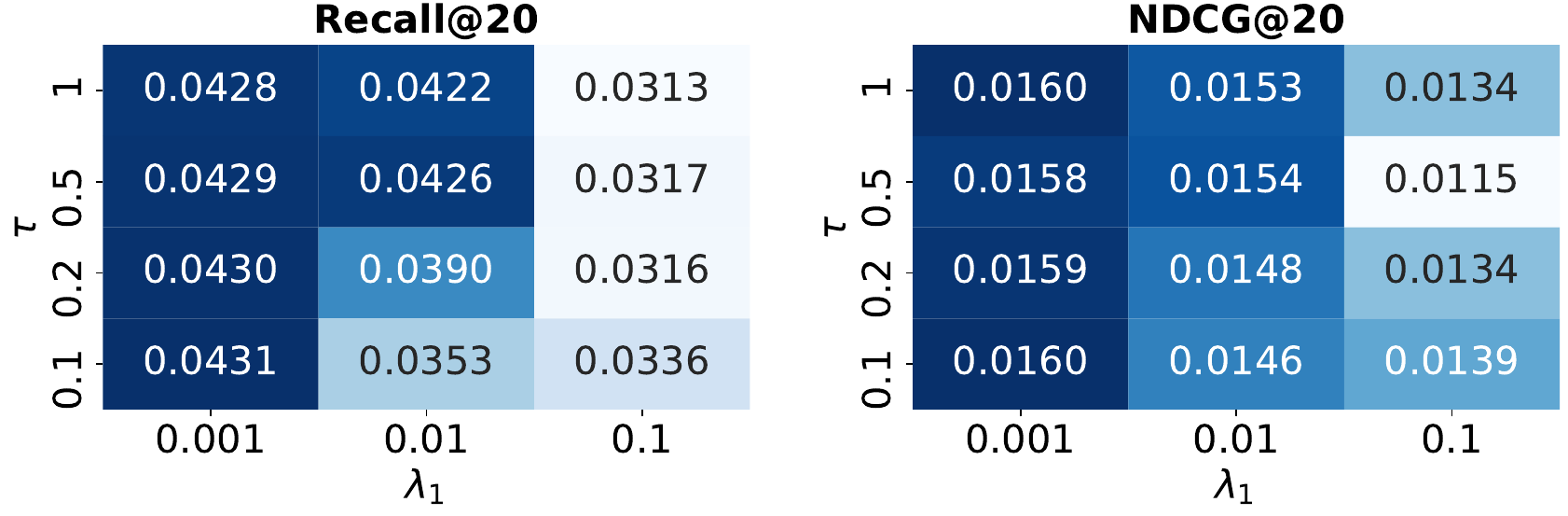}
        \textsf{\scriptsize \LL{(b) Allrecipes}}
    \end{minipage}
    
    \caption{{Performance comparison w.r.t different $\lambda_1$ and $\tau$.}}
    \label{Fig:CL}
\end{figure}

\subsubsection{Effect of the Latent Dimension $k_d$ in CD-Net}
The hyperparameter $k_d$ represents the dimension of the latent space, and the mapping from the item relationship space to the latent space makes the complexity of CD-Net controllable. \LL{As shown in Fig.~\ref{Fig:K}, on the Baby dataset (and similarly on the Allrecipes dataset, although not shown)}, both Recall@20 and NDCG@20 reach their maximum values at $k_d=2000$. In contrast, on the Sports and Clothing datasets, which involve a larger number of items, the metrics attain their maximum values at $k_d=3000$. These results suggest that larger item sets require a larger latent space dimension $k_d$ to prevent excessive compression of the item relationship information. However, excessively large values of \(k_d\) introduce additional noise, leading to performance degradation.

\subsubsection{Effects of $\lambda_1$ and $\tau$ in Representation Augmentation}
In the context of contrastive learning for representation augmentation, the contrastive learning loss weight $\lambda_1$ and the temperature coefficient $\tau$ are key hyperparameters. \LL{As shown in Fig.~\ref{Fig:CL}, the Baby dataset achieves the best performance with the hyperparameter combination $\{\lambda_1=0.1, \tau=0.2\}$, and although not shown in the figure, the other two Amazon datasets (Sports and Clothing) follow the same trend. For the Allrecipes dataset, the optimal performance is obtained  with the hyperparameter combination $\{\lambda_1 = 0.001, \tau = 0.1\}$.}


\subsubsection{Effects of $T$, $s$, and $\alpha_{\text{max}}$ in Diffusion Process}
\label{SEC:S}
\begin{figure}[tbp]
  \centering
  \includegraphics[width=\linewidth]{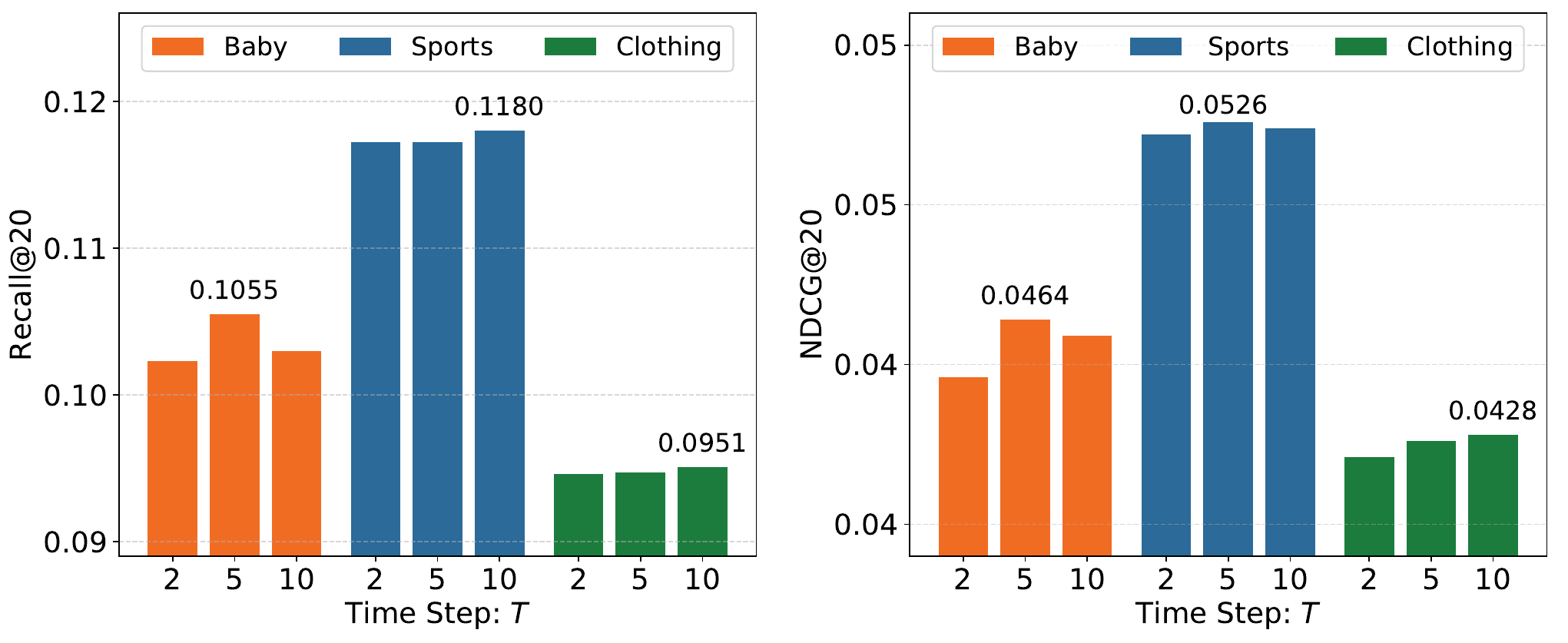}
  \caption{Performance comparison w.r.t different $T$.}
  \label{Fig:T}
\end{figure}

As shown in Fig.~\ref{Fig:T}, the time step $T$ has less impact on accuracy due to the lower noise levels. \LL{Notably, we empirically select $T=5$ for the Baby, Sports, and Allrecipes datasets and $T=10$ for the Clothing dataset to balance the performance and computation.} Table \ref{tab:s-alpha} demonstrates the effect of different values of $s$ and $\alpha_{\max}$ on the recommendation performance. \LL{Although the results on the Allrecipes dataset are not shown, it follows the same trend as Clothing.}


\begin{table}[t]
  \caption{Performance comparison w.r.t different $s$ and $\alpha_{\text{max}}$}
  \label{tab:s-alpha}
  \begin{threeparttable}
  \resizebox{0.47\textwidth}{!}{%
      \begin{tabular}{c|c|cc|cc|cc}
        \toprule
        \multicolumn{2}{c|}{{Dataset}} & \multicolumn{2}{c|}{{Baby}} & \multicolumn{2}{c|}{{Sports}} & \multicolumn{2}{c}{{Clothing}}\\
        \midrule[\heavyrulewidth]
        $s$ & $\alpha_{\text{max}}$ & R@20 & N@20 & R@20 & N@20 & R@20 & N@20 \\
        \midrule
        \multirow{2}{*}{0.002} & 0.02 & 0.1053 & 0.0460 & 0.1163 & 0.0519 & 0.0945 & 0.0422 \\
         & 0.05 & 0.1053 & 0.0458 & 0.1165 & 0.0517 & \textbf{0.0951} & \textbf{0.0428}\\
         \midrule
        \multirow{2}{*}{0.01} & 0.02 & \textbf{0.1055} & \textbf{0.0464} & \textbf{0.1172} & \textbf{0.0526} & 0.0931 & 0.0424 \\
         & 0.05 & 0.1051 & 0.0464 & 0.1165 & 0.0517 & 0.0940 & 0.0417 \\
        \bottomrule
      \end{tabular}  
  }
  \end{threeparttable}       
\end{table}

\section{Conclusion}
In this paper, we propose a novel Item Graph Diffusion for Multimodal Recommendation (IGDMRec), which leverages the conditional DM with classifier-free guidance to effectively optimize the structure of the item graph, while simultaneously enhancing recommendation accuracy through contrastive learning. To realize the denoising of the semantic item graph, IGDMRec proposes the BGD module, which takes the interaction data as conditioning information to guide the reconstruction of the semantic item graph, thereby generating the diffusion-aware item graph that fuses semantic and behavioral information. Meanwhile, a lightweight CD-Net is proposed to achieve denoising with manageable complexity. Furthermore, IGDMRec employs a contrastive representation augmentation scheme to fully utilize the diffusion-aware item graph and the semantic item graph, thereby enhancing the item representations. Extensive experiments demonstrate the superiority of IGDMRec compared to the state-of-the-art and the effectiveness of the key components in IGDMRec. 

\LL{While IGDMRec achieves consistent gains, its training still depends on explicit updates of the semantic item graph, which may incur additional computational overhead in large-scale scenarios. A promising direction is to explore new diffusion architectures that enable graph denoising without updating the full graph, thereby providing a more lightweight and scalable mechanism for graph refinement. In addition, leveraging diffusion models to better understand graph structures in RSs is a promising direction that may further advance graph-based recommendation methodologies.}

\ifCLASSOPTIONcompsoc
  \section*{Acknowledgments}
\else
  \section*{Acknowledgment}
\fi
This work was supported by the National Natural Science Foundation of China under Grants 62471357, 62372357, the Fundamental Research Funds for the Central Universities under Grant QTZX23072 and ZYTS241001.


\bibliographystyle{IEEEtran}
\bibliography{sample-base}

\end{document}